\documentclass[runningheads]{llncs}

\usepackage[T1]{fontenc}
\usepackage{graphicx}
\usepackage{hyperref}
\usepackage{siunitx}
\usepackage{microtype}
\usepackage{amsmath}
\usepackage{float}
\usepackage{booktabs}
\usepackage[english]{babel}

\usepackage{color}

\title{\emph{EnergyTwin}: A Multi-Agent System for Simulating and Coordinating Energy Microgrids}
\titlerunning{\emph{EnergyTwin}: Multi-Agent Simulation of Microgrids}

\author{Jakub Muszynski\inst{1}\orcidID{0009-0000-2797-6044} \and
Ignacy Walużenicz\inst{1}\orcidID{0009-0008-6269-362X} \and
Patryk Zan\inst{1}\orcidID{0009-0002-6772-8111} \and Zofia Wrona\inst{1}\orcidID{0009-0008-2863-5557} \and Maria Ganzha\inst{1}\orcidID{0000-0001-7714-4844} \and Marcin Paprzycki\inst{2,3}\orcidID{0000-0002-8069-2152} \and Costin~B\u{a}dic\u{a}\inst{4}\orcidID{0000-0001-8480-9867}}
\authorrunning{Muszynski et al.}
\institute{Warsaw University of Technology, Warsaw,
Poland,
\email{jakub.muszynski2.stud@pw.edu.pl},
\email{ignacy.waluzenicz.stud@pw.edu.pl}
\email{patryk.zan.stud@pw.edu.pl},
\email{zofia.wrona.dokt@pw.edu.pl}, \email{maria.ganzha@pw.edu.pl}\\
\and
Systems research Institute Polish Academy of Sciences, Warsaw, Poland\\
\and
University of Technology and Arts, Warsaw, Poland, \email{marcin.paprzycki@ibspan.waw.pl}\\
\and University of Craiova, Craiova, Romania, \email{costin.badica@edu.ucv.ro}}

\begin{document}
\maketitle

\begin{abstract}
Microgrids are deployed to reduce purchased grid energy, limit exposure to volatile tariffs, and ensure service continuity during disturbances. This requires coordinating heterogeneous distributed energy resources across multiple time scales and under variable conditions. Among existing tools, typically, power-system simulators capture physical behaviour but assume centralized control, while multi-agent frameworks model decentralized decision-making but represent energy with no physical grounding. In this context, the \emph{\emph{EnergyTwin}} is introduced, an agent-based microgrid simulation environment that couples physically grounded models with forecast-informed, rolling-horizon planning, and negotiations. Each asset is modeled as an agent, interacting with a central agent that obtains forecasts, formulates predictions, and allocates energy through contract-based interactions. \emph{EnergyTwin} targets tertiary-layer decision making and is extensible for digital-twin use. Its feasibility was evaluated in a university campus microgrid scenario where multiple planning strategies were compared. Achieved results show that forecast-driven rolling-horizon planning increases local energy self-sufficiency, maintains higher battery reserves, and reduces exposure to low-resilience operating states. They demonstrate also potential of \emph{EnergyTwin} as platform supporting research on resilient, negotiation-driven microgrids.
\end{abstract}

\section{Introduction} \label{sec:introduction}
Modern energy distribution systems evolve. Large consumers, e.g. university campuses, industrial parks, or research facilities are expected to meet sustainability targets (e.g., actual reductions in Scope~2 emissions \cite{sotos2015scope2} and higher on-site renewable penetration), limit exposure to volatile energy prices, and demonstrate continuity of service during power disturbances. Meeting these obligations typically involves: (i) increasing the share of local, low-carbon generation, e.g. via photovoltaic (PV) arrays, (ii) electrifying new end uses, e.g. transportation, and (iii) adding controllable flexibility in the form of battery storage and dispatchable loads. Microgrids -- electrically bounded subsystems that can coordinate these assets, operate independently from the upstream utility, making them a practical vehicle for achieving cost, carbon, and resilience objectives simultaneously.

Let us consider a university campus: a network of lecture halls, research laboratories with sensitive equipment, student dormitories, and athletic facilities. This merits introduction of microgrid, to meet sustainability goals and enhance resilience to power grid disturbances. Here, rooftops of campus buildings are fitted with PV arrays, a central battery energy storage system (BESS) is installed to store excess solar power, and a fleet of electric vehicle (EV) chargers is deployed. To maximize the energy self-sufficiency, this system must meet the operational objectives, e.g.: (1) minimize electricity costs, (2) reduce use of external energy, and (3) ensure that critical facilities (e.g. laboratories and data center) remain powered during upstream contingencies (e.g. feeder outages, voltage sags, etc., creating short-term instability or price spikes). This requires, among others, to dynamically (1) arbitrage time-of-use tariffs and shaving peaks, and (2) maximize on-site renewable energy self-consumption. This straightforward scenario encapsulates the complexity and promise of microgrids~\cite{lasseter2002microgrids,certs2002whitepaper}. 

Obviously, the ``campus energy manager'' must coordinate a heterogeneous collection of Distributed Energy Resources (DERs), each with own operational characteristics and constraints. Such coordination spans multiple, often conflicting, timescales. (1) Inverters connected to PVs and battery systems must react in milliseconds, to maintain local voltage and frequency stability~\cite{bahrani2024ibrlandscape}. (2) BESS charging and discharging schedule must be optimized over minutes and hours, to align with solar production forecasts and fluctuating electricity prices~\cite{hong2016stlf,antonanzas2016solar}. (3) EV charging must be managed to avoid overwhelming the local network during peak hours. (4) Sudden cloud cover, an unexpected grid voltage sag, or surge in demand (from a research experiment), demand immediate response.

This complex dynamics makes management of microgrids challenging. Here, physical experimentation may be infeasible, e.g. there is no simple way of de-energizing a laboratory to test the microgrid's islanding response. Therefore, high-fidelity simulation becomes a necessity. A credible simulation environment should allow stakeholders to explore different control strategies, stress-test the system under extreme weather or grid contingency scenarios, and quantify the economic and environmental benefits of investments, before installing the hardware. It forms the foundation for a digital twin, that can evolve with the physical system, providing a safe platform for operational planning and what-if analysis~\cite{aghazadeh2024dtses,kumari2023dtmicrogrid}.

Yet, building such simulator reveals fundamental tensions in current methodologies. Here, the traditional, optimization-based approaches model the system as a single, monolithic problem, to be solved by a central controller. While powerful, these methods can become brittle as the system scales and may need to accommodate diverse ownership models and privacy concerns inherent in real-world settings (e.g. a campus, where different departments manage their own assets). On the other side, the multi-agent systems (MAS) mirror the decentralized reality by treating individual entities (e.g. PV, BESS, etc.) as autonomous agents. This approach offers superior modularity and scalability. However, agent-based models often oversimplify the underlying electrical physics, potentially missing critical dynamics that may impact the stability and performance of the real system. Hence, a new class of simulation tools is needed.

To address this gap, the \emph{EnergyTwin}, a hybrid simulation platform designed to integrate physically grounded models with a structured, agent-based coordination layer, is introduced. The proposed system supports detailed representation of DERs, using validated components~\cite{king2004sandia,desoto2006pv,chen2006battery}, within a framework where agents negotiate and collaborate to manage the microgrid. By unifying the two paradigms, \emph{EnergyTwin} provides a realistic and extensible environment for designing and testing intelligent, decentralized control systems, required for the next generation of microgrids. In particular, this work contributes an open, modular architecture that captures the interplay between device physics and multi-actor decision-making, using the university campus as the guiding use case. The approach is experimentally validated on a realistic campus-scale microgrid. 

The remainder of this paper is organized as follows. Section~\ref{sec:sota} reviews the pertinent state of the art. Afterwards, Section~\ref{sec:method} describes the architecture of the \emph{EnergyTwin} system, while Section~\ref{sec:experiments} details the test cases and key performance indicators used for evaluation, reporting the obtained experimental results. Finally, Section~\ref{sec:discussion} analyzes the trade-offs of the proposed approach, whereas Section~\ref{sec:conclusion} summarizes the findings and outlines prospects of future work. 

It should be stressed that the code and the experimental scenarios are available to the community at:~\cite{EnergyTwin_github2025} and readers are invited to contribute to future development of the simulator.

\section{State of the Art}\label{sec:sota}
A microgrid is a geographically compact electrical subsystem that combines DERs, including PV generation, BESS, controllable loads, and flexible demand, with local control so that they can operate either connected to the main grid, or islanded (i.e. working independently of the main grid). Hence, the microgrid can be thought of as a small power system. It must keep voltage and frequency within bounds, share power among devices, and decide on improting or exporting energy at points of common coupling. From early demonstrations, these responsibilities have been organized into a layered, hierarchical control architecture separating fast device regulation from slower decisions at the network and market levels. In the \emph{primary} layer, devices stabilize voltage and frequency and share load, often through droop laws that modify a converter’s set-points as a function of measured electrical quantities. A \emph{secondary} layer restores nominal set-points and corrects residual errors after the primary layer work completion. The \emph{tertiary} layer examines economics and coordination with the upstream grid, over minutes to hours~\cite{guerrero2011hierarchical,olivares2014trends}. This separation aligns control tasks with their natural time scales, reduces complexity, and lets hardware and software evolve without compromising stability. This features was emphasized in recent and foundational surveys~\cite{li2023hierarchical,olivares2014trends}.

Early microgrids were built for reliability. Remote communities, hospitals, and research campuses that could not afford outages adopted local generation and storage to cope with disturbances. Today, they deliver also \emph{sustainability}, \emph{locality}, and \emph{independence}. Evidence from reviews and case studies shows that storage-enabled, renewable-rich microgrids can reduce emissions and distribution losses (\emph{sustainability}), supply local ancillary services (\emph{locality}), and maintain operation during faults and extreme events~\cite{uddin2023microgrids,ahmed2023review} (\emph{independence}). On the neighborhood or campus scale, they also support energy sharing among prosumers and \emph{transactive} coordination that aligns local environmental goals with market participation~\cite{machele2024interconnected}.

Most modern DERs interface through power electronic inverters. To denote such assets (PV, battery converters, some wind interfaces), the term \emph{inverter-based resources (IBRs)} is used. IBRs typically operate as \emph{grid-following} (synchronizing to an external voltage) or \emph{grid-forming} (synthesizing voltage or frequency references, useful in weak or islanded grids). Some existing works~\cite{bahrani2024ibrlandscape,esig2024gfm} discuss implications of these approaches for stability and operation in renewable-rich systems. However, as illustrated in the campus scenario (see, Section~\ref{sec:introduction}) no single viewpoint is sufficient: microgrid behaviour couples phenomena across wide time scales. With IBRs proliferating and DER diversifying, credible studies must analyze further (1) electrical physics (power flow and dynamics), (2) communication among devices and controllers, and (3) incentives created by tariffs or local rules.

Methodologies and tools have evolved accordingly. Time-domain methods and component models now better capture the dynamics of inverter-dominated systems \cite{lara2024revisiting}. Co-simulation frameworks such as \texttt{HELICS} (and \texttt{mosaik}) federate domain tools so power, communication, and market simulators can run together at scale~\cite{hardy2024helics,barbierato2022cosim}. In parallel, \emph{digital twins}, virtual replicas synchronized with field data, are used to explore designs, commission assets, and operate securely under uncertainty. Recent systematic reviews catalog enabling architectures and open challenges for energy systems~\cite{aghazadeh2024dtses,kumari2023dtmicrogrid}, illustrating that planning and operation now depend on simulation stacks that combine validated physical models, interoperable tooling, and scenario experimentation under uncertainty.

Multiple interesting microgrid work is related to optimization and control on networked physical models. Three pillars are especially worthy attention:

\begin{enumerate}

  \item \textbf{Optimal power flow (OPF).} The core of planning and steady-state operation. In its full AC form, OPF enforces power balance and equipment limits while optimizing an objective such as losses or cost. Convex relaxations can make large problems tractable. When systems grow or must respect privacy, distributed formulations, often built on the alternating direction method of multipliers (ADMM), decompose the network across feeders or actors, so that each solves a local problem and exchanges only limited information~\cite{erseghe2014admmopf}.

  \item \textbf{Model predictive control (MPC).} In receding-horizon decision making for DER coordination, MPC controller forecasts disturbances (e.g., irradiance, load), optimizes control actions in a moving time window, applies the first action, and repeats as new data arrives. Variants include economic MPC (explicit cost), stochastic/robust MPC (uncertainty), and hybrid MPC (mode switches such as on/off or grid-connected/islanded). These ideas span converter-level power quality to system-level energy management~\cite{hu2021mpcoverview,joshal2023mpc}.

  \item \textbf{Available tooling.} \texttt{OpenDSS} supports distribution-level quasi-static time series; \texttt{pandapower} brings power flow and OPF into Python workflows; \texttt{PyPSA} enables multi-period analysis~\cite{dugan2011opendss,thurner2018pandapower,brown2018pypsa}. Co-simulation (e.g., \texttt{HELICS}) connects power tools with market simulators for end-to-end studies~\cite{hardy2024helics,barbierato2022cosim}.

\end{enumerate}

When models and objectives are centrally curated, available methods produce strong baselines. However, as actors diversify (prosumers, aggregators), devices evolve, and rules change, monolithic assumptions become ineffective. The ``single optimization problem'' transforms into a community of independent decision makers with individual partial information. Here, microgrids become socio-technical systems with heterogeneous \emph{actors} (DER owners), facility managers, DSOs/aggregators (with local objectives, information, and constraints). Agent-based modeling (ABM) reflects this, by representing each entity as an \emph{agent} with state, capabilities, and interaction protocols (e.g., negotiation, contracts). In this case, distributed decisions become a feature rather than a problem. Moreover, privacy is respected, as only necessary signals are shared, and operational rules can evolve locally, without ``rewriting'' a central solver. Surveys document use of MAS across smart grids—protection, restoration, demand response, and market participation, highlighting strengths (modularity, scalability, adaptability) and challenges (verification, convergence guarantees)~\cite{izmirlioglu2024massurvey}. Case studies demonstrate real-time MAS energy management in low-voltage hybrid microgrids, where agents cooperate to meet system-level goals while preserving local optimization for devices and user preferences~\cite{elhafiane2024mas}. Historically, general agent platforms (e.g., \texttt{JADE}) and standard interaction protocols (e.g., the FIPA Contract Net) provided support for discovery, task allocation, and negotiation~\cite{bellifemine2007jade,fipa2002cnp}. Importantly, ABM complements, rather than replaces, classical optimization. An agent encapsulates an OPF or MPC and exposes only the signals required for coordination.

On the physics side, mature tools such as \texttt{OpenDSS}, \texttt{pandapower}, and \texttt{PyPSA} anchor power-flow and scheduling fidelity. Emerging open-source dynamic simulators extend electromagnetic-transient (EMT) analysis for grid-forming inverter-based resources, closing the gap for dynamics studies~\cite{su2024opensourcedynamic}. On the integration side, co-simulation frameworks (e.g. \texttt{HELICS} and \texttt{mosaik}) orchestrate cross-domain scenarios and were benchmarked for scalability when heterogeneous components interact~\cite{barbierato2022cosim,hardy2024helics}. On the agent side, two practical paths exist: (1) building microgrid-specific MAS models over general agent platforms (e.g., \texttt{JADE}) and couple them to power simulators, or (2) using agent-aware energy simulators, e.g., \texttt{GridLAB-D}, which embeds market/household agents alongside power-flow solvers~\cite{chassin2014gridlabd}.

The breadth of tooling reflects a common need: credible physics plus flexible decision architectures and interoperable experimentation. However, existing stacks tend to split along a seam: co-simulation frameworks (\texttt{HELICS}, \texttt{mosaik}) offer high-fidelity orchestration but do not involve agent negotiations, while agent-centric tools (\texttt{JADE}, \texttt{GridLAB-D}) ease multi-actor logic, but lack templates for tight synchronization with inverter-dominated dynamics, or repeatable agent–physics coupling. A simulator that unifies credible physics with reusable, protocol-driven agent coordination templates would address this issue directly.

Thus, to overcome the existing gap, \emph{EnergyTwin}, a hybrid microgrid simulator that integrates validated physical models and classical optimization with a multi-agent coordination layer was developed. Its design follows three principles.
\begin{enumerate}

  \item \textbf{Physical fidelity is a first-class design concern:} \emph{EnergyTwin} is structured to leverage established distribution-network and DER models rather than abstract commodity flow models.

  \item \textbf{Modular, agent-driven coordination:} agents encapsulate local MPC/OPF primitives and negotiate plans through lightweight, FIPA-style protocols that support plug-and-play DERs and stakeholder modularity.

  \item \textbf{Digital-twin extensibility:} the architecture is designed to synchronize with real telemetry so operators can rehearse scenarios, test control policies, and assess risk under uncertainty~\cite{aghazadeh2024dtses,kumari2023dtmicrogrid}.

\end{enumerate}

Here, agents representing DERs (PV, BESS), flexible loads, and an aggregator, coordinate over a rolling horizon using lightweight negotiation templates (FIPA-based) and simple, auditable planning steps. The architecture complements established physical modeling stacks and co-simulation settings, offering a modular path, combining classical optimization with agent-based coordination, as microgrids scale in complexity. This fills a gap in the state-of-the-art. It preserves the accuracy and interpretability of classical methods while adding flexibility, evolvability, and scalability of an agent-based systems approach, capturing how actual microgrids are planned and operated under uncertainty and change.

\section{System Overview and Methods}\label{sec:method}
\emph{EnergyTwin} leverages decentralized multi-agent framework to simulate coordinated operation of distributed energy resources, storage systems, and demand-side components, in a microgrid. The architecture is based on standard agent modeling principles and implemented using the JADE platform, with agents communicating via FIPA-compliant protocols. Each agent encapsulates a physical model and a decision logic, supporting autonomous, yet coordinated behavior. The resulting system allows detailed simulation of dynamic interactions among heterogeneous energy assets, and supports rolling-horizon planning, contract-based negotiation, and topic-based state synchronization. This section outlines agent roles, inter-agent communication, and the physical implications of their actions.
\subsection{Agent Architecture and Interactions} 
\begin{figure}[htp]
\centering
\includegraphics[width=0.75\textwidth]{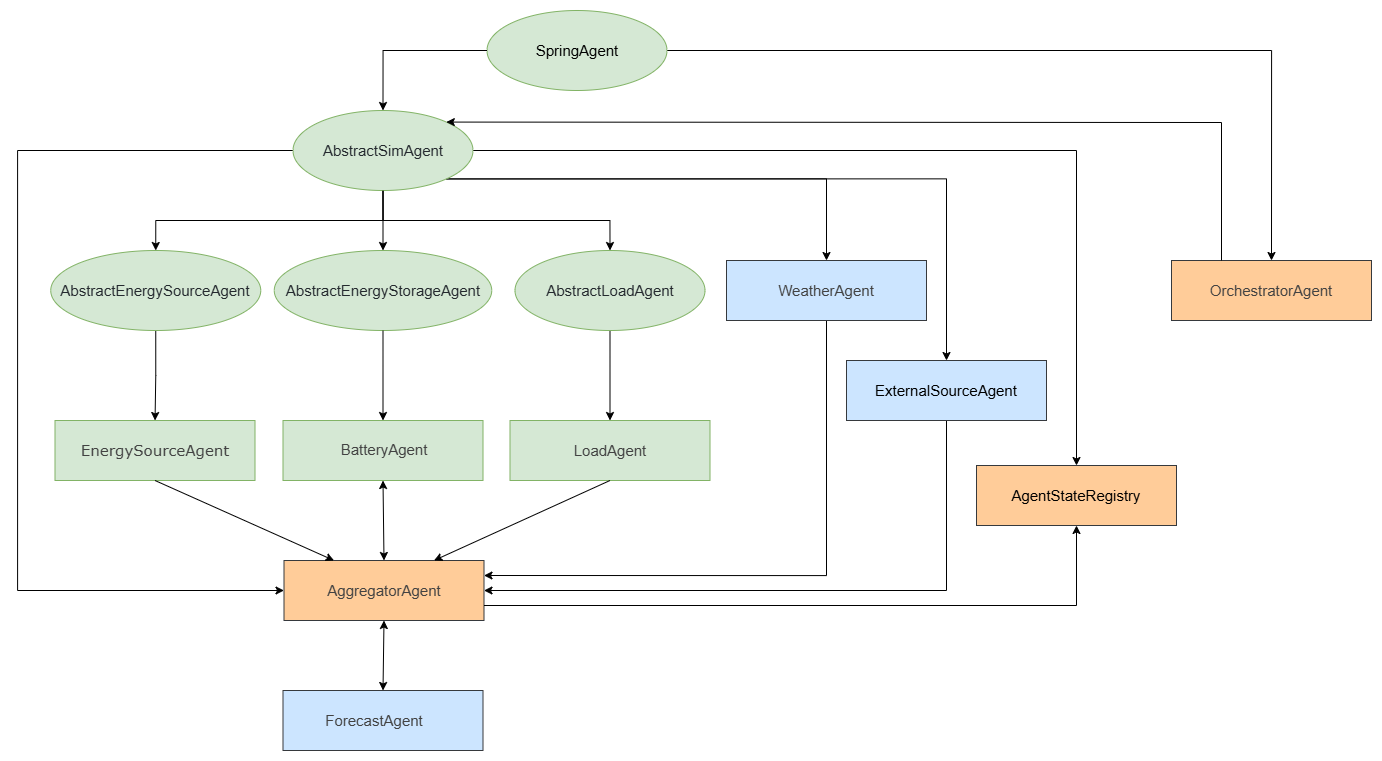}
\caption{Architecture of \emph{EnergyTwin}, showing agent types and primary interactions.}
\label{fig:system_agents}
\end{figure}

\emph{EnergyTwin} employs a hierarchical multi-agent architecture. In particular, the systems structure is illustrated in Figure~\ref{fig:system_agents}.

The central part of the system is the \texttt{OrchestratorAgent}. It is responsible for time synchronization and global coordination by maintaining consistent temporal reference across all agents, via periodic broadcast of \texttt{TickBroadcastBehaviour} messages. This ensures harmonized state updates, and ties agent activities to discrete simulation ticks. The state information of all participating agents is maintained by the \texttt{AgentStateRegistry}. The registry serves as a centralized repository enabling coherent state management and facilitateing information sharing, among others to the \texttt{AggregatorAgent}, which is responsible for the core planning function performed at each planning interval. 

Within the domain of energy management, agents inherit characteristics from particular types of agents based on their affiliation:
\begin{itemize} 
\item \texttt{AbstractEnergySourceAgent}: Agents responsible for generating energy.
\item \texttt{AbstractEnergyStorageAgent}: Agents responsible for storing energy.
\item \texttt{AbstractLoadAgent}: Agents responsible for modelling energy usage.
\end{itemize}

All above classes inherit from \texttt{AbstractSimAgent}, which introduces tick mechanisms and by itself inherits from \texttt{SpringAgent}, which encapsulates logic related to inter operation of Spring and JADE. Based on the affiliation classes, separate agent types representing individual components are instantiated: 
\begin{itemize} 
\item \texttt{EnergySourceAgent}: Represents distributed energy sources. It monitors energy generation capacity and participates in demand-response negotiations.
\item \texttt{AggregatorAgent}: Functions as an intermediary, consolidating requests and offers from subordinate agents within its purview. It orchestrates demand and supply balancing at a mesoscopic level. 
\item \texttt{LoadAgent}: Models discrete energy consumption. It submits energy demand requests and can adapt consumption behavior based on system signals. 
\item \texttt{BatteryAgent}: Emulates energy storage units with charging and discharging capabilities. It participates prominently in flexibility management by responding to grid shortfall or surplus situations. \item \texttt{ExternalSupplyAgent}: Represents (uncontrollable) external energy sources or markets, providing additional supply when local resources are insufficient. 
\item \texttt{WeatherAgent} and \texttt{ForecastAgent}: Supply exogenous data streams related to environmental conditions and predictive modeling, respectively, informing the decision-making of energy agents. 
\end{itemize} 

\emph{EnergyTwin} operates on discrete simulation ticks. The \texttt{OrchestratorAgent} advances the tick counter on a fixed schedule. At each tick, all agents are triggered (notification or topic message) to perform their cyclic behaviours (e.g. updating loads, computing bids). This global tick mechanism ensures that all agents progress synchronously. By waiting for each tick signal before continuing, the system avoids race conditions and maintains a consistent time-stepped simulation. 

\begin{figure}[htbp]
\centering
\includegraphics[width=0.65\textwidth]{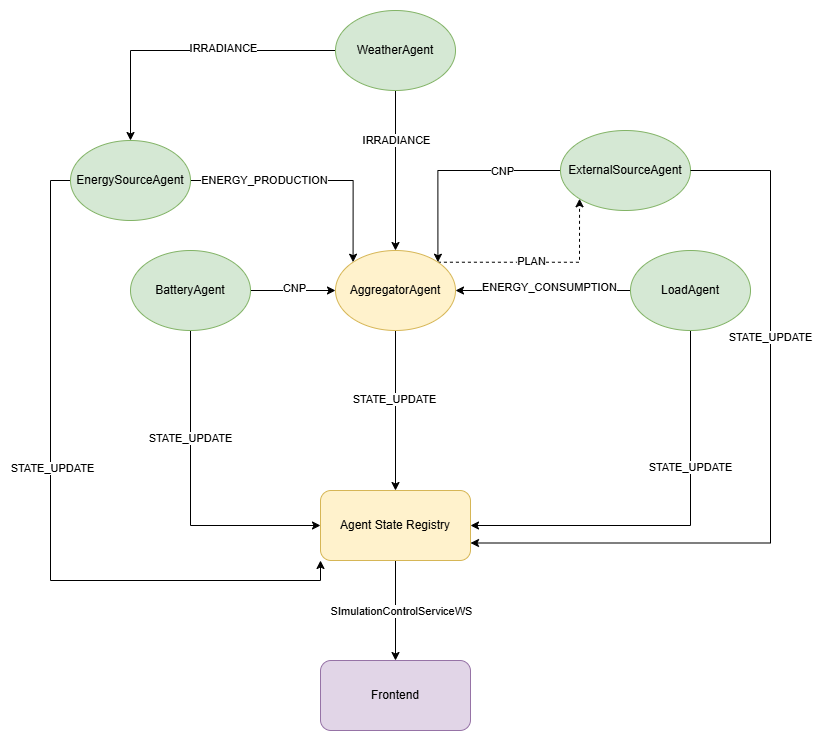}
\caption{JADE class hierarchy and communication flows in \emph{EnergyTwin}.}
\label{fig:interactions}
\end{figure}

Every agent publishes a \texttt{STATE\_UPDATE} message to the \texttt{AgentStateRegistry} at the end of its computation in a tick. Such messages contain agent's current metrics (generation, consumption, storage level, etc.). The registry then aggregates these updates into a single view of the grid’s state. This shared state can be queried by agents (e.g. \texttt{AggregatorAgent}) or pushed to an external interface, e.g. a frontend dashboard. By centralizing state information, the \texttt{AgentStateRegistry} decouples agents while enabling system-wide monitoring and coordination. Agent-to-agent communication in \emph{EnergyTwin} is built on FIPA-compliant ACL messages and principally uses the FIPA Contract Net Protocol (CNP) to structure scalable, distributed negotiations that resolve energy shortfalls and surpluses. Here, let us emphasize that in-depth exploration of negotiation mechanisms is not the main subject of this work, and will be further elaborated on in a separate work. For coordinated task allocation the \texttt{AggregatorAgent} aggregates local demands and acts as the CNP ``manager'' (using \texttt{ContractNetInitiator}), broadcasting \emph{call-for-proposals} (CFPs) upstream and collecting \texttt{PROPOSE} or \texttt{REFUSE} replies. Supplier agents such as the \texttt{BatteryAgent} and \texttt{ExternalSupplyAgent} implement the responder role (internal \texttt{BatteryCNPResponder} behavior and \texttt{ContractNetResponder}) and evaluate CFPs against current state and operational constraints to produce bids. Those bidding decisions directly influence charging/discharging actions and the physical flow of energy, e.g., leveraging stored energy to mitigate grid shortfalls or increasing charging to absorb surplus within operational limits. Then the \texttt{AggregatorAgent} evaluates proposals and awards contracts to the best candidates. It first gathers the current demand and supply states (via direct queries from agents or from \texttt{AgentStateRegistry}) and requests forecasts from the \texttt{ForecastAgent}. Using these inputs, it computes an energy dispatch plan over the next interval (for example, solving an economic dispatch or optimization problem). If the plan indicates insufficient or excessive local generation, the \texttt{AggregatorAgent} issues CFPs for additional energy from batteries or the grid. Thus, forecasting and negotiation are coupled: the agent uses predicted load/production to decide how much energy to procure through contracts. Communication proceeds hierarchically between aggregators and localized agents, with CFPs, responses, acceptances and execution messages sequencing through negotiation, acceptance and execution phases, enabling realistic emulation of complex energy system dynamics and supporting adaptive, decentralized management of generation, demand, storage and external inputs. In addition to point-to-point ACL exchanges, agents also use topic-based channels for system-wide signals. For example, the \texttt{OrchestratorAgent} publish tick events or global updates on shared topics to which agents subscribe, simplifying broadcast announcements and allowing simultaneous notification of many agents while preserving the scalability. All agents' interactions are outlined in Figure~\ref{fig:interactions}.

Apart from representing the energy system components, a credible microgrid simulation study requires two more elements: (1) physically grounded models of the assets and their environment, and (2) an intelligent, forward-looking coordination process that plans how those assets should behave. Accordingly, the following sections details the models and methods implemented in \emph{EnergyTwin} to satisfy these requirements. Let us first focus on physical component models.

\subsection{Physical Component Simulation}

\begin{figure}[htp]
  \centering
  \includegraphics[width=0.6\linewidth]{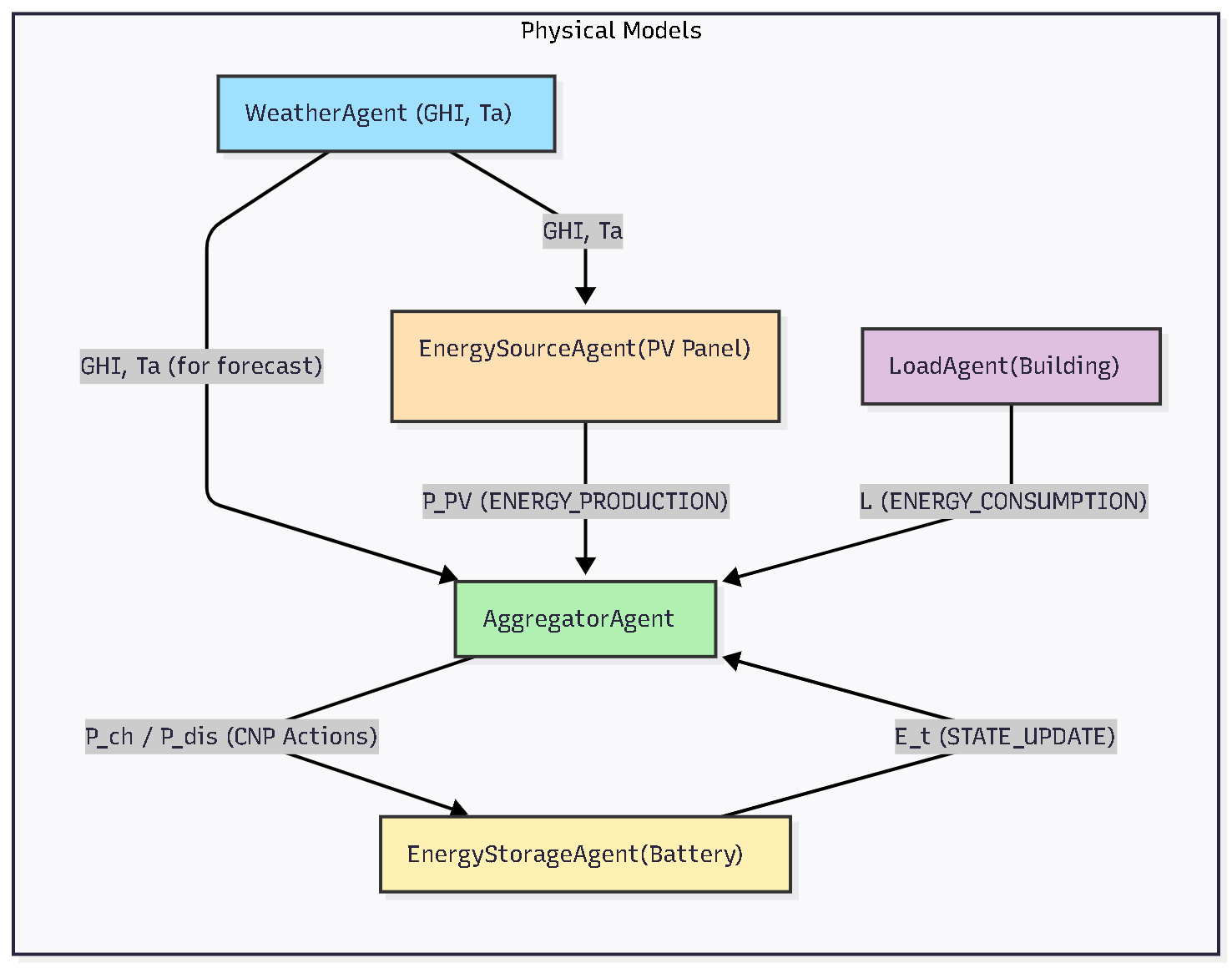}
  \caption{Physical modelling data flow in \emph{EnergyTwin} system.}
  \label{fig:physical-models}
\end{figure}

Each physical asset is represented by an agent that encapsulates its operational model and constraints. These agents are driven by environmental parameters sent by \texttt{WeatherAgent}. Figure~\ref{fig:physical-models} shows the data flow within the components of the \emph{EnergyTwin}. 

For the purpose of the simulation, several physical models were implemented, to support accurate modeling of real grid conditions.

\begin{enumerate}

\item \textit{Weather and Environmental Model} The \texttt{WeatherAgent} is responsible for generating the primary exogenous drivers: global horizontal irradiance (GHI) in \si{W/m^2} and ambient air temperature ($T_a$) in \si{\celsius}. Here, configurable, first-order model based on sinusoidal functions for the diurnal solar and temperature cycles is used. The model, configured with parameters \texttt{sunriseTick}, \texttt{gPeak}, \texttt{tempMeanDay}, and \texttt{tempMeanNight}, provides a lightweight, predictable, and controllable environment for testing control logic. Stochastic weather variability is introduced by adding Gaussian noise (scaled by \texttt{sigmaG} and \texttt{sigmaT}) at each time step. To convert GHI to plane-of-array (POA) irradiance, well established transposition and clear-sky models~\cite{perez1987diffuse,ineichen2002linke,holmgren2018pvlib} are used.

\item \textit{Photovoltaic Generation Model} The \texttt{EnergySourceAgent} models the power output of a PV array. This agent implements a well-established thermal model to capture the dependency of performance on the weather~\cite{king2004sandia,desoto2006pv}. First, the cell temperature ($T_c$) is calculated from the ambient temperature ($T_a$) and the incident irradiance ($G$), using a standard Nominal Operating Cell Temperature (NOCT) model, shown in Eq.~\ref{eq:noct}:
\begin{equation}
\label{eq:noct}
T_c = T_a + (G / G_{\text{NOCT}}) \times (T_{\text{NOCT}} - T_{\text{amb,NOCT}})
\end{equation}
where $T_c$ is the resulting cell temperature, $T_a$ is the current ambient air temperature, and $G$ is the incident irradiance on the panel. Parameters: $T_{\text{NOCT}}$ (the cell temperature at NOCT conditions), $T_{\text{amb,NOCT}}$ (the ambient temperature at which NOCT is defined, typically \SI{20}{\celsius}), and $G_{\text{NOCT}}$ (the irradiance at which NOCT is defined, typically \SI{800}{W/m^2}), are configurable.

Next, the reference efficiency at the standard test conditions (STC; $\eta_{\text{STC}}$), is adjusted using a negative temperature coefficient, $\gamma$, to find the true operating efficiency, $\eta$, as shown in Eq.~\ref{eq:pv-eff}:
\begin{equation}
\label{eq:pv-eff}
\eta = \eta_{\text{STC}} \times (1.0 + \gamma \times (T_c - T_{\text{STC}}))
\end{equation}
where $\eta$ is the final operating efficiency, $\eta_{\text{STC}}$ is the module's rated efficiency and $\gamma$ is the temperature coefficient (negative value indicate performance loss related to heat). $T_c$ is the cell temperature from Eq.~\ref{eq:noct}, and $T_{\text{STC}}$ is the standard test condition temperature, defined as \SI{25}{\celsius}. In model, the simulated PV output realistically degrades during hot, high-irradiance conditions.

\item \textit{Energy Storage Model} The \texttt{EnergyStorageAgent} implements a discrete-time state-of-charge (SOC) model~\cite{chen2006battery,plett2015bms}. The agent tracks the stored energy ($E_t$) in \si{kWh}, governed by the state update equation:
\begin{equation}
\label{eq:soc}
E_{t+1} = E_t + (\eta_c P_{ch,t} \Delta t) - (\frac{1}{\eta_d} P_{dis,t} \Delta t) - (E_{\text{self}} \Delta t)
\end{equation}
where $E_{t+1}$ is the energy stored in the battery at the \emph{next} time step, and $E_t$ is the energy at the \emph{current} time step. The change in energy is determined by: $\eta_c$, charging efficiency, $P_{ch,t}$, charging power applied during the interval, $\eta_d$, discharging efficiency, $P_{dis,t}$, discharging power drawn during the interval, and $\Delta t$, duration of time step. Note that $P_{dis,t}$ is divided by its efficiency to model that more energy must be \emph{drawn} from the battery than is delivered to the load. Finally, $E_{\text{self}}$ accounts for small, continuous energy loss due to self-discharge. All actions are constrained by battery's nominal energy capacity, its C-rate (power limits), and operational SOC bounds.

\item \textit{Demand Simulation Model} The \texttt{LoadAgent} simulates electrical demand using a synthetic, profile-based approach. The used model mimics the behaviour of a university campus. It combines a deterministic daily baseline with stochastic variability. The baseline is defined by a 24-hour schedule (e.g., 80\% of nominal load during working hours 08--18, 50\% during low-working periods 06--08 and 18--20, and 10\% at night). At each time step, a random delta is applied to this baseline to simulate unpredictable, high-frequency fluctuations.

\end{enumerate}

\subsection{Forecasting Methodology}~\label{sec:method-forecast}

To enable predictive control, the \texttt{AggregatorAgent} must move beyond myopic, rule-based dispatch and anticipate future conditions. This requires a robust, accurate, and computationally efficient forecasting pipeline. Several forecasting paradigms have been evaluated to assess their suitability. As a result, a \textbf{Random Forest (RF) regressor}~\cite{breiman2001rf} was selected as the primary forecasting engine. It delivers a high accuracy, while being robust to overfitting, and computationally efficient for both training and inference. This balance is critical for an online system that must retrain during operation and generate new forecasts rapidly.

\begin{figure}[htbp]
  \centering
  \includegraphics[width=0.85\linewidth]{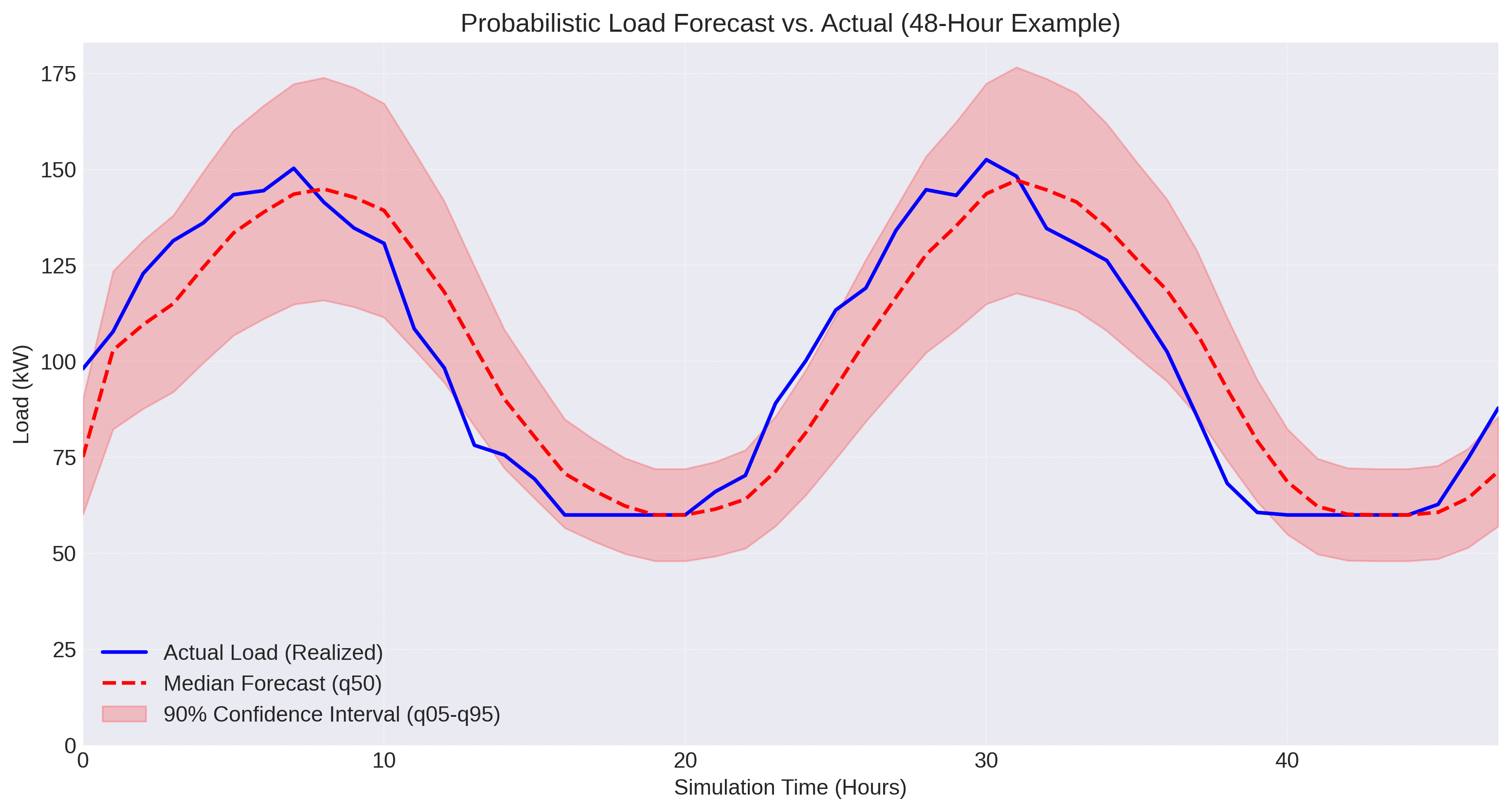}
  \caption{Example forecasting via RandomForest model in a 48-hour time horizon, with 7-day history sample for training.}
  \label{fig:forecast-example}
\end{figure}

The forecasting component (\texttt{ProbabilisticForecaster} service) uses the \texttt{smile} machine learning library. Instead of using raw time-series values, it is trained on a feature set derived from historical data, including: lagged values, moving averages, trends, exogenous environmental data, and, critically, cyclical temporal features to represent time-of-day and weekly patterns. The key advantage of the ensemble nature of the RF lies in ability to provide probabilistic forecasts, not just a single point prediction~\cite{hong2016stlf,antonanzas2016solar}. By analyzing the distribution of the predictions from individual trees in the forest, the service returns the \textbf{low (q05), median (q50), and high (q95) quintiles} for future load and PV generation (see, Figure~\ref{fig:forecast-example}), forming the basis of the uncertainty-aware planning.
\vspace{-7mm}

\subsection{Optimization and Scenario-Based Planning}
The final step is to translate the forecasts into an optimal and physically feasible dispatch plan. The system is designed to support full stochastic optimization but, currently, implements computationally lighter, deterministic-equivalent control.

The quintile forecasts are passed to the \texttt{QuantileTreeGenerator} service, which constructs discrete three-branch scenario tree that provides a compact representation of future uncertainty. It generates three distinct scenarios, each with an assigned probability: a low-quintile scenario (pessimistic: high load, low PV), a median scenario (median load, median PV), and a high-quantile scenario (optimistic: low load, high PV). This structure is designed to be fed, in the future, into a stochastic optimization solver.

Currently, \texttt{AggregatorAgent} invokes a deterministic planner for predictive-planning. This service formulates a \textbf{Linear Programming (LP)} problem, based \emph{only} on the \textbf{median (q50) scenario}. The planner, which uses the Google OR-Tools \texttt{GLOP} solver~\cite{ortools2024}, minimizes cost function comprising total grid energy import and a small proxy for battery degradation, to account for the cycling costs. The optimization is subject to two key sets of constraints: device constraints (e.g. SOC dynamics in Eq.~\ref{eq:soc}) and power balance constraint. The power balance must hold for every time step $k$ in the planning horizon, as shown in Eq.~\ref{eq:power-balance}:
\begin{equation}
\label{eq:power-balance}
P_{PV,k} + \sum_{i} P_{dis,i,k} + P_{grid,k}^{\text{imp}} = L_{k} + \sum_{i} P_{ch,i,k} + P_{grid,k}^{\text{exp}}
\end{equation}

This equation ensures that all energy sources (on the left) equal all uses of energy (on the right). Specifically: $P_{PV,k}$ is the forecasted solar power at time $k$. $P_{dis,i,k}$ is the discharge power from battery $i$, summed over all available batteries. $P_{grid,k}^{\text{imp}}$ is the power imported from the external grid. These must balance: $L_{k}$, the forecasted load at time $k$; $P_{ch,i,k}$, the charge power for battery $i$, summed over all batteries; and $P_{grid,k}^{\text{exp}}$, any surplus power exported to the grid.

\begin{figure}[htp]
  \centering
  \includegraphics[width=0.5\linewidth]{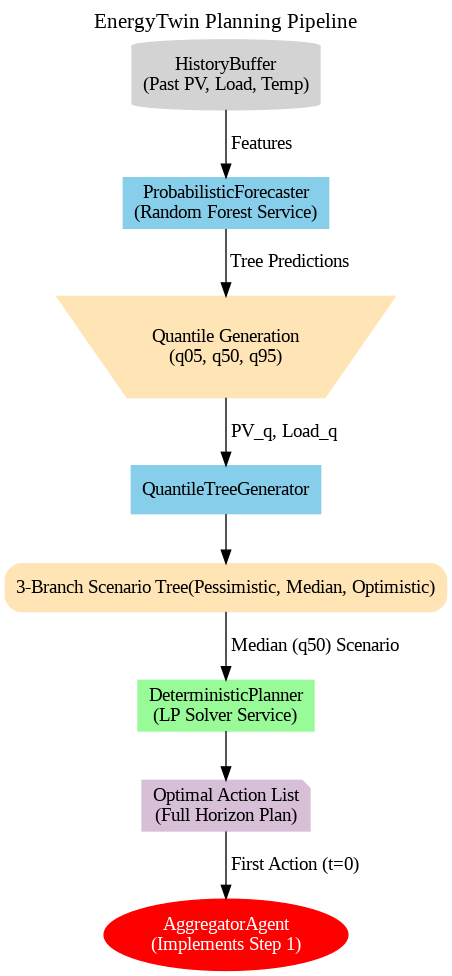}
  \caption{\emph{EnergyTwin} pipeline used to plan the behaviour of microgrid in future timesteps.}
  \label{fig:planning-pipeline}
\end{figure}

The planner solves the LP and returns an optimal sequence of \texttt{Action} objects, which describe amount of energy to charge, discharge and import, for the entire horizon. Next, \texttt{AggregatorAgent} implements only \emph{first} action in the sequence. This forecast-and-plan process is then repeated at the next planning interval, creating a \textbf{rolling-horizon execution}, similar to Model Predictive Control, that continuously adapts to new information. Full pipeline is outlined in Figure~\ref{fig:planning-pipeline}.

\subsubsection{Baseline Myopic Control}~\label{sec:method-phys}
Along the predictive planning mode, for comparison, the system supports a \textbf{baseline myopic control mode}, which operates without forecasting. Instead, at each time step, the \texttt{AggregatorAgent} observes the current, realized net load (i.e., $L_t - P_{PV,t}$). Then issues commands to \texttt{EnergyStorageAgent} and \texttt{ExternalSupply} instances to meet the immediate deficit or absorb the surplus based on a fixed, deterministic priority list. For example, in the dispatch priority, surplus PV is first used to charge the battery (before any export). Conversely, deficits are first covered by battery discharge (before importing from the external source). This non-predictive, rule-based approach serves as the baseline, against which the operational gains, of the predictive optimization, are measured.

\section{Experimental Evaluation}\label{sec:experiments}
The experimental study evaluates the operational impact of introducing forecasting and rolling-horizon planning into the coordination loop. Two controller modes are compared:
\begin{itemize}
    \item \textbf{Baseline mode} -- a myopic controller that reacts greedily, at each tick, using only instantaneous measurements as described in Section~\ref{sec:method-phys}.
    \item \textbf{Predictive mode} -- a planning controller that, at every decision instant: (i) requests short-horizon forecasts of load and photovoltaic (PV) production, (ii) solves linear programming problem over finite horizon, and (iii) applies the first resulting action, analogous to the receding-horizon or to the Model Predictive Control strategy (see, Section~\ref{sec:method-forecast}).
\end{itemize}

Both modes run on the same physical microgrid configuration, under identical weather and load realizations. Planning in predictive mode is activated at tick \(168\) (introducing a ``warm-up period''), after which the controller continuously updates its plan every two ticks with a forecast horizon of four ticks.

\subsection{Microgrid Scenario}\label{sec:exp-scenario}
The simulated microgrid represents a prosumer-style campus site with local renewable generation, on-site battery storage, and a non-interruptible load. The specific configuration, used in experiments, is as follows:
\begin{itemize}
    \item \textbf{Weather model.} The \texttt{WeatherAgent} provides irradiance and ambient temperature, using a diurnal profile (sunrise at tick~6, sunset at tick~18, peak irradiance at tick~12 with \(\texttt{gPeak} = 1000 \, \text{W}/\text{m}^2\)) with stochastic perturbations to emulate cloud cover and short-timescale variability.
    \item \textbf{Photovoltaic source.} A PV array (\texttt{PVMain}) composed of \(200\) panels (area \(2.0\,\mathrm{m}^2\) each, nominal efficiency \(0.2\), temperature coefficient \(-0.004\,\mathrm{K}^{-1}\)), modeled using irradiance- and temperature-dependent conversion efficiency (see, Section~\ref{sec:method-phys}).
    \item \textbf{Battery energy storage.} A battery (\texttt{MainBattery}) of capacity \(100\) kWh, charge/discharge efficiency \(0.95\), C-rate~1.0, and self-discharge \(10^{-3}\) per tick. State of charge (SoC) is tracked with the discrete-time model from Eq.~\ref{eq:soc}.
    \item \textbf{Load.} A building-scale consumer (\texttt{CampusBuilding}) follows a campus daily profile with daytime peaks and reduced night demand, scaled by a nominal load factor of \(50.0\), equivalent to a maximum demand of 50kW.
    \item \textbf{External source.} An ``infinite'' external supplier with fixed unit cost and capacity limit \(100\) kW, representing a distribution grid connection.
\end{itemize}
The experiment runs for \(336\) ticks, where one tick corresponds to one hour. All trends and metrics were reported in tick time (hours since the start of simulation). To assess the effect of predictive planning, several physically interpretable metrics were extracted directly from the logs of the simulation.
\paragraph{Cumulative Energy Balance Ratio (CEBR).}
CEBR measures how much energy the system produced relative to how much it consumed up to time \(t\). Formally,
\begin{equation}
\label{eq:exp-cebr}
\mathrm{CEBR}_t = 100 \times \frac{E^{\mathrm{prod}}_t}{E^{\mathrm{cons}}_t}\,,
\end{equation}
where \(E^{\mathrm{prod}}_t\) and \(E^{\mathrm{cons}}_t\) are the cumulative locally produced and consumed energy, respectively, up to tick \(t\). Values above \(100\%\) indicate that local production covers total demand and creates potential surplus. Values below \(100\%\) indicate persistent reliance on the external source. Figure~\ref{fig:exp-cebr} plots \(\mathrm{CEBR}_t\) for both controller modes over the full run. Furthermore, the \emph{incremental} energy balance ratio (IEBR) is computed after the activation of predictive planning. IEBR is the per-tick analogue of CEBR, defined as
\begin{equation}
\label{eq:exp-iebr}
\mathrm{IEBR}_t = 100 \times
\frac{\Delta E^{\mathrm{prod}}_t}{\Delta E^{\mathrm{cons}}_t}, \quad
\Delta x_t := x_t - x_{t-1}.
\end{equation}
It is evaluated only for ticks with \(\Delta E^{\mathrm{cons}}_t>0\). The post-activation mean IEBR reflects how effectively the controller balances the net load \emph{during} operation, as opposed to CEBR, which considers the entire past. Values above 100\% indicate that local sources not only met the demand but also generated surplus energy (e.g. to eb stored in the BSS).

\paragraph{Battery reserve behaviour.}
Battery SoC is a direct proxy for operational flexibility, since high SoC increases the ability to cover the next demand spike or an outage of the PV source. Figure~\ref{fig:exp-soc} shows SoC trajectories for both modes. Two derived indicators are tracked:
\begin{itemize}
    \item \textbf{Average post-activation SoC.} The arithmetic mean of \(\mathrm{SoC}_t\) for ticks \(t > 168\). High values indicate that the controller is actively maintaining headroom for resilience.
    \item \textbf{Battery Reserve Index (BRI\(_{\geq 50\%}\)).} The percentage of post-activation ticks where \(\mathrm{SoC}_t \geq 50\%\). This measures how often the battery is kept in a ``comfortable'' reserve band, stipulated as ``at least half full''.
\end{itemize}

The \textit{complementary failure mode} is captured by \textbf{scarcity proxy}: the fraction of post-activation ticks where PV generation is zero (night-time or heavy clouds) \emph{and} \(\mathrm{SoC}_t < 5\%\). This indicates an imminent risk of unmet demand if the external source was to be degraded, because neither PV nor storage could sustain the load. Both BRI\(_{\geq 50\%}\) and the scarcity proxy are depicted in Figure~\ref{fig:exp-kpis}.

Finally, total battery cycling is quantified through the number of equivalent full cycles (EFC). EFC is computed from SoC excursions as:
\begin{equation}
\label{eq:exp-efc}
\mathrm{EFC} = \frac{1}{200}
\sum_{t=2}^{T} \left|\mathrm{SoC}_t - \mathrm{SoC}_{t-1}\right|.
\end{equation}
This approximates how many full charge-discharge cycles the battery executed during the run. Higher EFC reflects more active energy shifting, and corresponds to increased wear. The denominator 200 normalizes total SoC excursion into equivalent full cycles, assuming that a ``full cycle'' corresponds to a 100\% $\rightarrow$ 0\% $\rightarrow$ 100\% swing (i.e., 200 percentage points of cumulative SoC change).

\paragraph{Forecasting accuracy.}
Predictive mode relies on short-horizon forecasts of load and PV output, generated by the forecasting service (Section~\ref{sec:method-phys}), which uses a Random Forest regressor~\cite{breiman2001rf}. The quality of those forecasts is evaluated at each planning instant after activation. Forecast error is defined as the absolute difference between predicted and realized values (in kW) for both load and PV. Mean absolute error (MAE) is obtained as the average of those absolute errors, over all planning instants. Figure~\ref{fig:exp-forecast} shows the distribution of absolute errors over time, together with the corresponding MAE for load and PV.

\subsection{Quantitative Results}\label{sec:exp-results}
\begin{figure}[htp]
    \centering
    \includegraphics[width=0.8\linewidth]{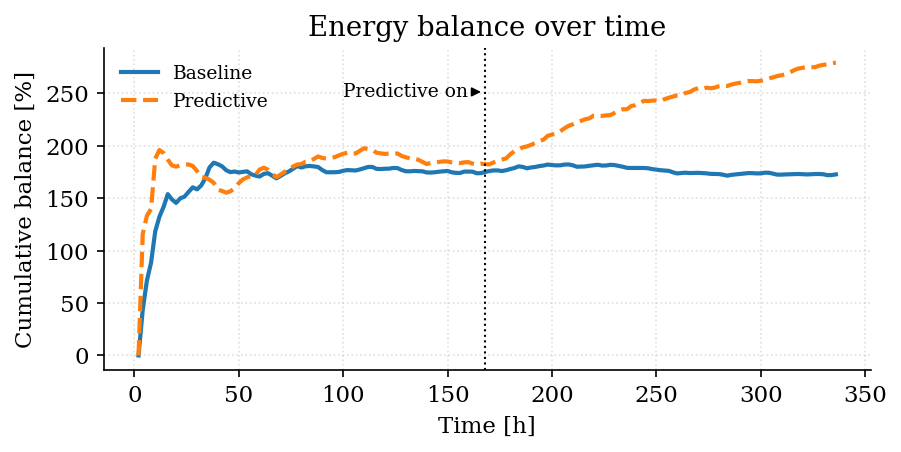}
    \caption{A timeseries of Cumulative Energy Balance Ratio (CEBR) across simulation.}
    \label{fig:exp-cebr}
\end{figure}

Let us first describe the results of the experiments obtained when focusing on the energy balance. Figure~\ref{fig:exp-cebr} shows that the cumulative energy balance under predictive mode diverges upward after tick~168, i.e.\ once the forecasting-enabled planning begins. The final cumulative ratio reaches \(279.13\%\) in the predictive mode versus \(172.59\%\) for the baseline control (Table~\ref{tab:metrics-summary}). This indicates that predictive planning substantially increases the share of locally supplied energy and thus reduces the effective reliance on external sources.

The superiority of the predictive controller becomes more apparent when examining the \emph{rate} of improvement. After predictive planning is enabled (at tick~168), IEBR (in the predictive mode) stabilizes at consistently higher values than in the baseline case. Thus, short-horizon planning actively reshapes the dispatch of the battery and the PV output so that each new demand interval is met locally. The effect is also captured quantitatively in Table~\ref{tab:metrics-summary}. Here, the mean post-activation IEBR reaches \(502.09\%\) under predictive control, compared to \(214.53\%\) for the baseline, which suggests that when forecasting and the rolling-horizon optimization are applied, the microgrid is covering each new slice of demand from local resources, rather than falling back to the external source. This behaviour directly complements the increase in cumulative balance observed in Figure~\ref{fig:exp-cebr}.
\begin{figure}[htp]
    \centering
    \includegraphics[width=0.8\linewidth]{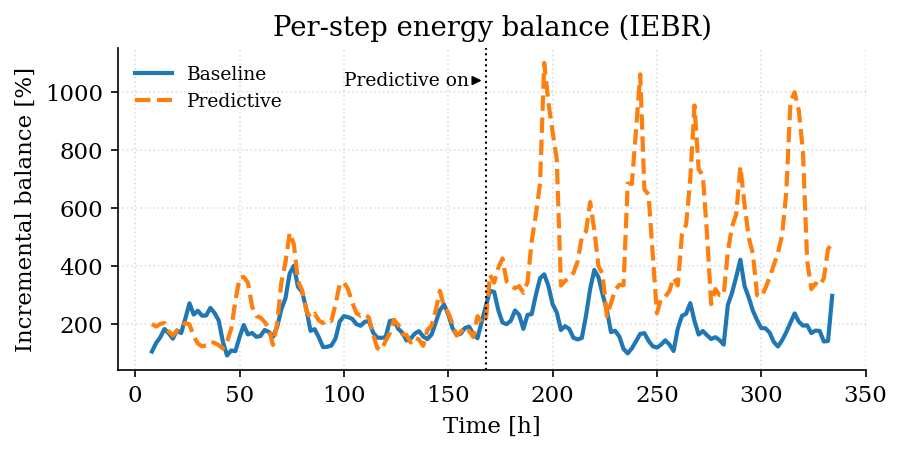}
    \caption{Incremental Energy Balance Ratio (IEBR) over simulation time.}
    \label{fig:exp-iebr}
\end{figure}

\subsubsection{Storage utilisation and operational resilience}
\begin{figure}[htp]
    \centering
    \includegraphics[width=0.9\linewidth]{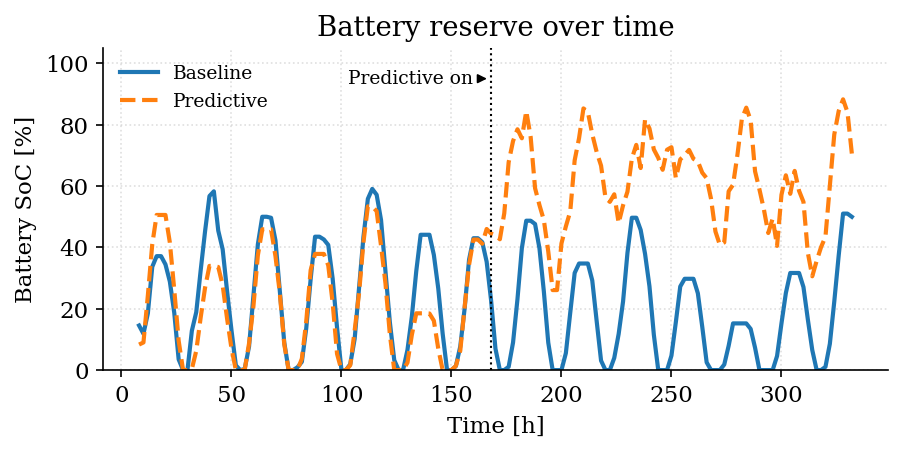}
    \caption{Battery state of charge (SoC) over simulation time.}
    \label{fig:exp-soc}
\end{figure}

\begin{figure}[htp]
    \centering
    \includegraphics[width=0.9\linewidth]{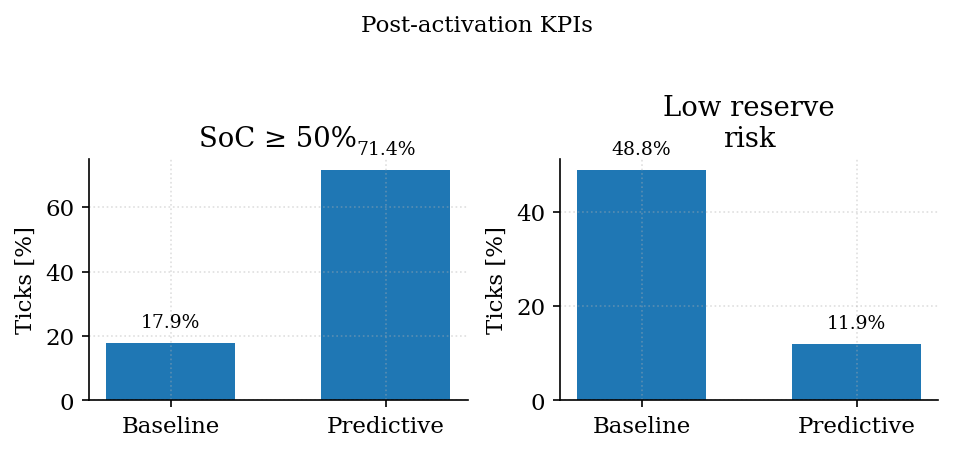}
    \caption{Post-activation key performance indicators derived from the battery trajectory illustrated by percentage of post-activation ticks during which (1, left) SoC is at least \(50\%\) (BRI), (2, right) SoC below \(5\%\) while PV generation is zero.}
    \label{fig:exp-kpis}
\end{figure}

Battery behaviour clearly differs between the two modes. As shown in Figure~\ref{fig:exp-soc}, after the predictive controller becomes active, the battery is maintained at substantially higher SoC levels and rarely drops near empty (overnight). Quantitatively, the mean post-activation SoC increases from \(18.63\%\) (baseline) to \(61.76\%\) (predictive), and the battery remains above \(50\%\) SoC for \(71.43\%\) of post-activation ticks, compared to only \(17.86\%\) in the baseline (Table~\ref{tab:metrics-summary}). This is also summarized in Figure~\ref{fig:exp-kpis} (left panel).

The scarcity proxy in Table~\ref{tab:metrics-summary} further highlights the system resilience. Under baseline control, almost half of post-activation ticks (\(48.81\%\)) fall into a ``low reserve'' state (PV production is zero and the battery is nearly depleted; \(<5\%\) SoC). Predictive control reduces this to \(11.90\%\), see right panel of Figure~\ref{fig:exp-kpis}. Note that in a microgrid, low-reserve means ``high vulnerability'', as failure of external sources would immediately force curtailment, or load shedding. The predictive controller accumulates charge, ahead of night-time, or low-generation, periods.

The improved reserve profile comes at the cost of more active cycling. The number of equivalent full cycles increases from \(12.58\) in baseline mode to \(20.75\) under predictive planning (Table~\ref{tab:metrics-summary}). This means that predictive control not only ``holds'' the battery at a safe SoC, but it also continuously shifts energy to protect the microgrid against later scarcity.

\subsubsection{Forecasting quality}
\begin{figure}[htp]
    \centering
    \includegraphics[width=0.9\linewidth]{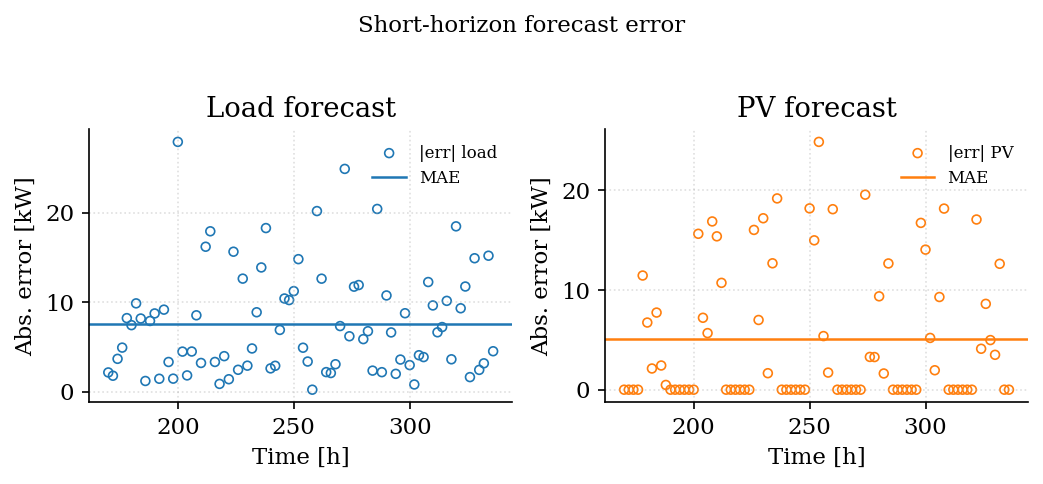}
    \caption{Short-horizon forecast error for load (left) and PV generation (right) over the predictive planning window. Dots indicate absolute error at each planning instant. The horizontal line is MAE of the forecasting model.}
    \label{fig:exp-forecast}
\end{figure}

The predictive controller uses short-horizon forecasts of load and PV output, to plan charging and discharging. Figure~\ref{fig:exp-forecast} shows the absolute forecast error (in kW) at each planning instant after activation, along with the corresponding MAE. 
The forecast MAE is \(7.56\,\mathrm{kW}\) for load and \(5.06\,\mathrm{kW}\) for PV (Table~\ref{tab:forecast-summary}). Overall, errors are bounded and exhibit no runaway drift. This shows that the forecasting pipeline is sufficiently accurate to support rolling-horizon planning, even for non-stationary operating conditions, induced by the controller. 
This closes the loop between perception (forecasting), negotiations and planning (\texttt{AggregatorAgent}), and actuation (battery and source agents).

\subsubsection{Resilience under unplanned disturbances} \label{sec:exp-stress}
Next, controller behaviour was evaluated for two standard stress scenarios: (i) loss of PV source (\emph{PV outage}) and (ii) abrupt multi-hour increase in demand (\emph{load spike}). Both scenarios were executed using the same configuration, described in Section~\ref{sec:exp-scenario}. The disturbances were injected only after predictive planning started (post tick~168). All reported metrics are as introduced in Section~\ref{sec:exp-scenario}.

\begin{figure}[htp]
    \centering
    \includegraphics[width=0.8\linewidth]{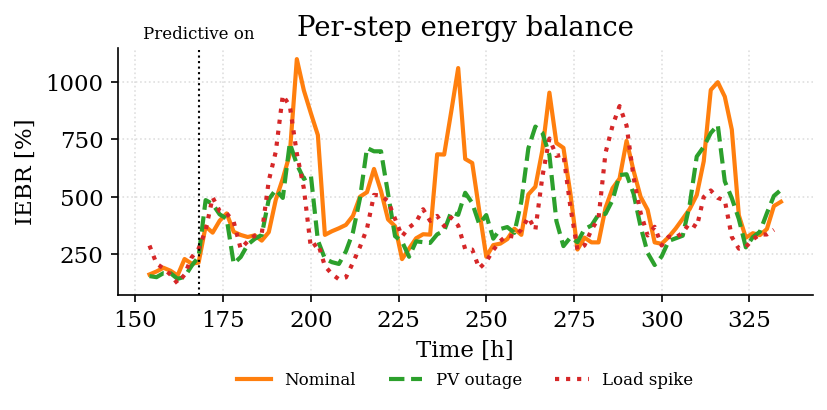}
    \caption{Per-step IEBR in predictive mode under nominal operation, PV outage (loss of local generation), and load spike (sustained high demand), for predictive planning.}
    \label{fig:exp-iebr-stress}
\end{figure}

To evaluate how the predictive planning responds dynamically to the contingencies, Figure~\ref{fig:exp-iebr-stress} plots the IEBR (Eq.~\ref{eq:exp-iebr}), for three predictive-mode trajectories: nominal operation, PV outage, and load spike. IEBR captures, for each tick, how much demand was supplied locally (from PV and storage), instead of being imported from the outside. Values above \(100\%\) indicate that local energy capacity (production and storage discharge) exceeded local consumption (i.e.\ represent net export potential).

The PV outage trajectory in Figure~\ref{fig:exp-iebr-stress} closely tracks the nominal predictive trajectory (after tick 169). It can be seen that after the main PV source is disabled, the per-step local coverage of demand does not collapse immediately. This is consistent with the aggregate indicators. The final CEBR, for PV outage, remains at \(269.27\%\) (compared to \(279.13\%\) for nominal predictive and \(172.59\%\) for baseline control), and the mean post-activation IEBR remains elevated, at \(440.03\%\), well above the baseline controller, at \(214.53\%\). Thus, for PV outage, the predictive planner draws from, previously stored, battery energy, instead of immediately using the external source. This is reflected also in storage-related resilience metrics: the post-activation average state of charge reaches \(65.02\%\), and the BRI\(_{\geq 50\%}\) is \(72.62\%\), meaning the battery remains at or above \(50\%\) SoC for nearly three quarters of post-activation ticks. The scarcity proxy -- the fraction of post-activation ticks with \(\mathrm{SoC}<5\%\) and zero PV -- drops to \(7.14\%\), which is lower than even the nominal predictive run (\(11.90\%\)) and far below the baseline controller (\(48.81\%\)). This suggests that the predictive planner maintains headroom specifically to tolerate the generation loss.

The load spike scenario has qualitatively different results. Figure~\ref{fig:exp-iebr-stress} shows that immediately after the surge in demand, IEBR remains high. Here, the predictive controller attempts to satisfy the excess demand locally by aggressively discharging BSS. However, when the high demand persists, the load spike trajectory begins to deviate (downward) from the nominal and the PV outage trajectories, particularly in the latter ticks. This corresponds to the battery becoming energy-limited; i.e., post-activation average SoC drops to \(10.57\%\), the Battery Reserve Index BRI\(_{\geq 50\%}\) collapses to \(8.43\%\) (from \(71.43\%\) in nominal predictive case), and the scarcity proxy rises to \(48.19\%\), matching the baseline controller's \(48.81\%\). Thus, under sustained over-consumption the planner initially buffers the demand, but the battery is eventually depleted (as in the baseline, non-predictive, case). This is also reflected in the per-step balance: although the mean post-activation IEBR remains high in aggregate (\(414.13\%\)), Figure~\ref{fig:exp-iebr-stress} shows that, obviously, it cannot be held once storage is depleted.

These results illustrate two distinct resilience behaviours. For a supply-side contingency (PV outage), predictive planning keeps the battery sufficiently charged in advance, and local service quality and per-step local coverage remain similar to the nominal predictive operation. For a demand-side contingency (multi-hour load spike), predictive planning initially shields the system by discharging storage, but prolonged stress pushes it into a vulnerable low-reserve state. Then, the scarcity risk resembles that of the myopic baseline controller. This suggests that extending the planner with an explicit reserve margin or risk-aware constraint may be a natural next step. The \texttt{AggregatorAgent} could enforce a minimum survivability budget for extreme load events, rather than spending all stored energy to maximise short-term self-supply. However, such decision has to consider, for instance, likelihood and importance of extreme events, vis-a-vis the cost.

\subsection{Summary of Quantitative Indicators} \label{sec:exp-summary}
Table~\ref{tab:metrics-summary} summarizes the most relevant post-activation indicators. Overall, predictive planning raises the average state of charge, increases the availability of stored energy, and substantially reduces time spent in critical low-reserve states, while improving both the cumulative and the per-step energy balance of the microgrid. 
\begin{table}[htp]
\centering
\caption{Post-activation performance indicators. CEBR (Eq.~\ref{eq:exp-cebr}). IEBR (Eq.~\ref{eq:exp-iebr}). BRI. Scarcity proxy: ticks with SoC below \(5\%\) and zero PV generation. EFC (Eq.~\ref{eq:exp-efc}).}
\label{tab:metrics-summary}
\begin{tabular}{lrr}
\toprule
Metric & Baseline & Predictive \\
\midrule
Final CEBR [\%] & 172.59 & 279.13 \\
Mean IEBR post-activation [\%] & 214.53 & 502.09 \\
Avg SoC post-activation [\%] & 18.63 & 61.76 \\
BRI $\geq 50\%$ post-activation [\% of ticks] & 17.86 & 71.43 \\
ScarcityProxy post-activation [\% of ticks] & 48.81 & 11.90 \\
Equivalent full cycles [--] & 12.58 & 20.75 \\
\bottomrule
\end{tabular}
\end{table}

Table~\ref{tab:forecast-summary} reports the MAE for load and PV generation forecasts, confirming that the forecasting subsystem provides reliable short-horizon estimates.
\begin{table}[htp]
\centering
\caption{Short-horizon forecast accuracy in predictive mode. MAE in kW, computed over all planning instants after activation.}
\label{tab:forecast-summary}
\begin{tabular}{l r}
\toprule
Metric & Predictive model \\
\midrule
Load forecast MAE [kW] & 7.56 \\
PV forecast MAE [kW] & 5.06 \\
\bottomrule
\end{tabular}
\end{table}

Taken together, these results show that enabling forecasting and rolling-horizon planning in the \texttt{AggregatorAgent} substantially improves energy self-sufficiency, raises usable reserves in storage, and reduces exposure to low-resilience operating states, while relying on forecast models whose error levels remain within the range required for effective short-horizon scheduling.

\section{Discussion of experimental results} \label{sec:discussion}
The experimental study, reported in Section~\ref{sec:experiments}, evaluates a multi-agent microgrid coordination framework under two control regimes: purely reactive, instantaneous dispatch versus forecast-informed, rolling-horizon planning embedded in the \texttt{AggregatorAgent}. The resulting trajectories differ not only in aggregate energy self-sufficiency, but also in storage use, exposure to scarcity, and behaviour under stress events, such as PV loss and prolonged load surges. Let us interpret those outcomes from three perspectives. (1) Operational implications of forecast-driven planning, with attention to autonomy and resilience. (2) Developed architecture usability, seen in terms of modularity and degree of decentralization, situating it within the landscape of hierarchical and multi-agent microgrid control. (3) Current scope and limitations, clarifying which aspects of real microgrid operation have been captured by \emph{EnergyTwin} and which remain out of scope or simplified.

\subsection{Implications of Forecast-Driven, Multi-agent Coordination}

The experimental results indicate that embedding short-horizon forecasting and rolling-horizon optimization into a multi-agent coordination loop changes how a microgrid behaves. After predictive planning is activated (at tick~168), the microgrid transitions from a purely reactive system to one that actively stages energy in anticipation of the future. This shift is visible in three dimensions.

(I) Energy self-sufficiency improves substantially. The CEBR climbs to \(279.13\%\) under predictive planning, compared to \(172.59\%\) for the myopic baseline, and the IEBR after activation averages \(502.09\%\) versus \(214.53\%\) for the baseline. Figures~\ref{fig:exp-cebr} and \ref{fig:exp-iebr} show that the predictive controller serves new demands from local PV and storage rather than importing energy from the external source. This is consistent with established benefits of Model Predictive Control (MPC) and receding-horizon scheduling in microgrids, where limited foresight may be enough to unlock substantial gains in economic performance and/or resilience~\cite{hu2021mpcoverview,joshal2023mpc}.

(II) Storage is treated as a strategic reserve. Under predictive planning, the average post-activation state of charge (SoC) is \(61.76\%\) and the battery remains above \(50\%\) SoC for \(71.43\%\) of all post-activation ticks. The myopic baseline, in contrast, maintains an average SoC of only \(18.63\%\) and keeps SoC above \(50\%\) for just \(17.86\%\) of ticks. Figure~\ref{fig:exp-soc} and the left panel of Figure~\ref{fig:exp-kpis} show that the predictive controller regularly preserves headroom before night-time or low-irradiance periods, thus reducing exposure to ``low reserve'' situations (SoC $<5\%$ and zero PV) from \(48.81\%\) in the baseline to \(11.90\%\) under predictive planning. In microgrid terms, the system remains usable longer, even if the external source is degraded, aligning with the reliability and survivability objectives that motivated microgrids in campuses, hospitals, and remote communities~\cite{olivares2014trends,uddin2023microgrids,ahmed2023review}.

(III) Stress-testing confirms that these behaviours are not brittle. When local PV is disabled in the \emph{PV outage} scenario, the per-step IEBR does not collapse but, for a while, closely tracks nominal predictive performance. Post-activation SoC actually increases to \(65.02\%\) on average, the Battery Reserve Index (SoC $\geq 50\%$) reaches \(72.62\%\), and the scarcity proxy falls to \(7.14\%\), indicating that the controller staged sufficient charge to tolerate a (temporary) loss of generation. Thus, forward-looking planning implicitly creates resilience for supply-side disturbances without requiring an explicit ``contingency mode''.

The \emph{load spike} scenario represents a complementary failure mode. The predictive controller initially absorbs a multi-hour demand surge by aggressively discharging storage, maintaining high IEBR, similar to the nominal operation. Over time, however, the sustained spike depletes the battery. IEBR, for the load spike, gradually drifts downward as the battery exhausts. This behaviour shows that the current planning strategy prioritizes immediate local self-supply, even at the cost of entering a future low-reserve state. In a real-world deployment, that policy would be unacceptable for critical loads. It also exposes a limitation of strictly cost-oriented dispatch, and motivates the inclusion of explicit survivability constraints, or reserve margins, in the planning objective.

Overall, these results demonstrate that even with relatively simple components the coordination logic substantially improves autonomy and resilience. This is achieved without requiring a detailed, centralized day-ahead market optimization. Instead, forecasting and planning are embedded in an \emph{multi-agent} negotiation loop. This pattern is compatible with the socio-technical reality of many campus-scale microgrids, where assets are locally owned and managed but still expected to participate in a supervisory energy management scheme~\cite{guerrero2011hierarchical,izmirlioglu2024massurvey,elhafiane2024mas}.

\subsection{Modularity, Hierarchy, and Degree of Decentralization}

\emph{EnergyTwin} was motivated by an identified research gap (Section~\ref{sec:sota}), which is addressed by the tool's architecture in two ways.

(A) Each Distributed Energy Resource (DER) is encapsulated in a dedicated agent with a realistic physical model. \texttt{EnergySourceAgent} applies irradiance- and temperature-dependent PV conversion efficiency. \texttt{EnergyStorageAgent} enforces SoC dynamics with charge/discharge efficiency and self-discharge. \texttt{LoadAgent} reproduces realistic campus-style demand profile with stochastic variation. These models can be easily simulated, but capture the dominant first-order behaviours, relevant for short-horizon scheduling: PV derates when hot, batteries lose round-trip energy, and load is not flat. The shared \texttt{AgentStateRegistry} and synchronized ticks from \texttt{OrchestratorAgent} maintain a coherent global state, without forcing controllers to run in the same codebase.

(B) The economic and operational decision logic is encapsulated in an \texttt{AggregatorAgent} that both (i) solves an optimization problem and (ii) negotiates with other agents via FIPA CNP. As a result, (a) \texttt{AggregatorAgent} is undeniably a central coordinator: it requests forecasts, runs an LP, and issues awards. From an optimization perspective, the architecture therefore resembles a classic hierarchical microgrid controller with a supervisory tertiary layer \cite{guerrero2011hierarchical,olivares2014trends} rather than a fully distributed ADMM style OPF \cite{erseghe2014admmopf}. At the same time, (b) use of explicit messaging, rather than shared memory or function calls, means that each DER (agent) can, in principle, refuse, counter-propose, or price its flexibility. The contract mechanism becomes the natural place to inject local objectives, privacy constraints, or ownership rules without rewriting the global solver.

This hybrid design reflects how actual microgrids are often governed. Campuses, industrial parks, and community energy districts frequently operate under an internal “aggregator” (energy manager) that is responsible for cost, compliance, and resilience, while individual assets (e.g. lab buildings, data centers, EV fleets) have local autonomy, safety constraints, and private priorities. \emph{EnergyTwin} captures this situation. Moreover, the architectural choice makes it straightforward to evolve toward more distributed formulations. For example, multiple \texttt{AggregatorAgent} instances could be introduced, each running a local OPF or MPC, and interacting via price signals or ADMM-like exchanges~\cite{erseghe2014admmopf}, or DER agents could be extended to submit bids that include not only energy availability but also degradation cost, user comfort constraints, or emissions targets. Hence, \emph{EnergyTwin} bridges classical centralized scheduling and agent-based coordination.

\subsection{Scope of Validity and Limitations}

Despite the gains, reported in Section~\ref{sec:experiments}, several simplifying assumptions constrain the present scope.

\textit{Network and dynamics fidelity.} The physical layer is modeled as a single-bus microgrid with hourly timesteps. Line limits, voltage regulation, and converter dynamics are not captured. Moreover, grid-forming versus grid-following behaviour of inverter-based resources, and related small-signal stability, in low-inertia or islanded regimes, are outside the current simulation. Thus, the current \emph{EnergyTwin} targets tertiary-level scheduling and resilience assessment (minutes to hours) rather than primary/secondary voltage and frequency control (sub-second to seconds)~\cite{bahrani2024ibrlandscape,esig2024gfm,lara2024revisiting}. Integrating a distribution power flow model (e.g., via \texttt{OpenDSS} or \texttt{pandapower}) and  EMT-scale inverter dynamics would allow \emph{EnergyTwin} to capture feeder congestion, reactive power support, and grid-forming stability under islanding, aligning it with the full hierarchical stack discussed in Section~\ref{sec:sota}.

\textit{Economic realism.} Currently, the optimization objective minimizes imported energy and penalizes battery cycling via a simple proxy. Tariff structures, demand charges, export compensation, carbon pricing, and contractual penalties for non-delivery are simplified or omitted. Likewise, grid-code and regulatory constraints -- for example, minimum reserve requirements, maximum export limits at the point of common coupling, or ride-through obligations -- are not enforced. This is a deliberate simplification: it isolates the effect of forecast-driven planning on the physical behaviour. Future work should integrate realistic tariff models and feeder constraints so that the economic conclusions (e.g., cost savings, peak shaving, arbitrage value) can be represented quantitatively and explicitly.

\textit{Uncertainty handling and resilience.} Although the forecasting subsystem produces quantile trajectories (low/median/high) and constructs a scenario tree, the current planner solves a deterministic linear program using only the median scenario. The experimental results demonstrate that this median-only strategy is sufficient to pre-charge storage for predictable scarcity (night, low irradiance), to gracefully absorb a PV outage. However, the same strategy can exhaust reserves under a prolonged load spike, pushing the system back into the high-risk regime seen in the baseline control. This suggests that explicitly encoding survivability constraints, reserve margins, or robust/stochastic objectives into \texttt{AggregatorAgent}'s optimizations is necessary. This would allow the planner to rationally trade short-term energy self-sufficiency against long-term resilience.

\textit{Timescale separation.} The system uses hourly ticks and a four-tick forecast horizon. Real microgrids operate across nested timescales, i.e., milliseconds for inverter control, seconds to minutes for secondary control and load shedding, minutes to hours for economic dispatch and demand response~\cite{guerrero2011hierarchical,olivares2014trends,hu2021mpcoverview}. Hence, \emph{EnergyTwin} occupies only the slower end of this hierarchy. This is appropriate for evaluating forecasting, negotiation, and rolling-horizon planning, but fast dynamics and protection behaviour (e.g., anti-islanding, fault ride-through, safety interlocks) remain out of scope.

Taken together, these limitations define the present \emph{EnergyTwin} implementation as a tertiary-layer, campus-scale coordination sandbox. It focuses on rolling-horizon dispatch, forecasting under uncertainty, and negotiation among heterogeneous actors, evaluated through resilience-oriented metrics. The architecture, however, is explicitly structured to admit higher-fidelity electrical models, richer economic rules, and more distributed coordination schemes. Addressing these extensions is the subject of future work and is discussed in Section~\ref{sec:conclusion}.

\section{Conclusion and Future Work}\label{sec:conclusion}
The rapid growth of inverter-based distributed energy resources, combined with increasingly heterogeneous ownership and operational objectives, continues to push microgrids beyond the assumptions of monolithic, centrally optimized control. At the same time, practical deployments such as university campuses, industrial parks, and research facilities demand tools that not only capture physical constraints with sufficient fidelity but also reflect the institutional reality of negotiation, limited information sharing, and resilience requirements under uncertainty. The \emph{EnergyTwin} platform addresses this need by coupling interpretable physical models, short-horizon forecasting, and rolling-horizon optimization within a structured multi-agent architecture.

On the modelling side, each Distributed Energy Resource is represented as an autonomous software agent with its own operational limits and physical behaviour, e.g. photovoltaic generation depends on irradiance and panel temperature, storage evolves according to explicit state-of-charge dynamics with efficiency and self-discharge, and building demand follows a realistic, stochastic campus profile. A dedicated \texttt{WeatherAgent} and \texttt{ForecastAgent} supply exogenous drivers and learned short-horizon forecasts, while an \texttt{OrchestratorAgent} and \texttt{AgentStateRegistry} provide a synchronized view of the system state. This organization is consistent with how microgrids are commonly conceptualized in hierarchical control literature, where fast local loops stabilize hardware and slower supervisory layers perform dispatch and coordination~\cite{guerrero2011hierarchical,olivares2014trends}.

Considering coordination, \emph{EnergyTwin} instantiates an \texttt{AggregatorAgent} with two roles. First, it solves a linear programming problem over a receding horizon, analogous to economic Model Predictive Control, using median forecast trajectories to generate a cost-aware dispatch plan. Second, it interacts with other agents through FIPA-style negotiation patterns, issuing calls for proposal to storage and external suppliers and awarding ``contracts'' for energy provision. This design preserves a clear supervisory authority -- appropriate, for example, for a campus energy manager -- while maintaining modularity, privacy boundaries, and the potential for agents to internalize local constraints or preferences \cite{izmirlioglu2024massurvey,elhafiane2024mas}.

Quantitative evaluation on a campus-like microgrid scenario demonstrates that this approach materially changes operational outcomes. After activating predictive planning, the system (i) increases its local energy coverage, with final CEBR rising from \(172.59\%\) in the myopic baseline to \(279.13\%\); (ii) maintains substantially higher battery reserves, with SoC above \(50\%\) for \(71.43\%\) of post-activation ticks instead of \(17.86\%\); and (iii) reduces time spent in critically scarce operating states (night-time, SoC $<5\%$, no PV) from \(48.81\%\) of post-activation ticks to \(11.90\%\). These improvements persist even under a simulated PV outage, indicating that proactive energy staging, guided by forecasts, can deliver meaningful resilience to supply-side shocks. At the same time, a sustained load spike exposes the fact that the planner willingly depletes storage to maintain short-term autonomy, eventually driving the system back into a vulnerable state. This behaviour highlights that pure cost minimization is not sufficient for resilience, and that survivability-aware planning constraints are needed.

Taken together, these results support three overarching conclusions.

\begin{enumerate}
    \item Forecast-informed, rolling-horizon planning, even over a short horizon and with lightweight models, can meaningfully improve both autonomy and resilience of a microgrid. This is observable in energy balance metrics, battery reserve posture, and robustness to PV loss.

    \item Embedding a planner inside an agent-based coordination layer enables a modular, extensible architecture that mirrors how real microgrids are governed, i.e., assets remain autonomous, with physical constraints and private objectives, yet supervisory actor can negotiate and enforce system-level goals.

    \item The multi-agent structure exposes issues with classical dispatch logic. The load spike experiment makes shows that achieving economic efficiency and maximizing local self-consumption must be balanced against survivability objectives, reserve policies, and contractual obligations for critical loads.
\end{enumerate}

These conclusions point directly to future development steps. First, the planning layer will be extended from a deterministic median-scenario LP to an explicitly risk-aware, or stochastic, optimizer. The existing quantile scenario tree already provides low/median/high trajectories of load and PV. Integrating those trajectories into the objective and constraints will allow \texttt{AggregatorAgent} to maintain minimum reserves against infrequent but high-impact disturbances (such as multi-hour demand surges), rather than spending all stored energy to improve short-term self-sufficiency. This direction includes enforcing explicit survivability margins and reserve state-of-charge limits as first-class constraints, rather than as an emergent behaviour.

Second, the negotiation layer will be generalized beyond a single agent. Multi-aggregator or peer-to-peer configurations would allow groups of DERs (for example, independent campus buildings or co-owned assets in a local energy community) to coordinate through price signals, rather than through a single supervisory allocator. Incorporating distributed optimal power flow primitives and ADMM-like exchanges~\cite{erseghe2014admmopf} will make it possible to study privacy-preserving coordination, where no actor is forced to reveal its full internal model.

Third, the physical layer could be deepened. Coupling \emph{EnergyTwin} to existing distribution-system simulators, e.g., \texttt{OpenDSS} or \texttt{pandapower}, could introduce voltage, current, and protection constraints, enabling studies of feeder congestion, reactive power support, and compliance with grid codes. Incorporating inverter-level dynamics and grid-forming behaviour could extend the time resolution toward primary and secondary control layers~\cite{bahrani2024ibrlandscape,esig2024gfm,lara2024revisiting}. This would make \emph{EnergyTwin} a realistic digital twin, where a live microgrid can be stress-tested under contingencies, cybersecurity events, and evolving tariff structures.

Finally, validation on real telemetry and tariff data from an operational microgrid remains an essential milestone. Such validation would make it possible to quantify cost savings, export/import patterns, emissions impact, and compliance with reliability requirements, and would allow direct benchmarking against established campus energy management practices.

In summary, \emph{EnergyTwin} provides an open, extensible foundation for studying microgrid coordination as both a physical and an institutional problem. By unifying forecast-driven planning, structured agent negotiation, and interpretable physical models, it offers a path toward microgrid digital twins that are not only electrically credible but also decision-aware, resilience-aware, and capable of evolving with policy and ownership structures.

\section*{Acknowledgements}
\textbf{Funding.} Work of Maria Ganzha and Marcin Paprzycki has been supported in part by the CETPartnership, Clean Energy Transition Partnership under the 2023 joint call for research proposals, co-funded by the European Commission (GA N°101069750) and with the funding organizations detailed at 
\url{https://cetpartnership.eu/funding-agencies-and-call-modules}. Moreover, this work is also a result of bilateral cooperation between the Polish Academy of Sciences and Academia Română; project: ''Intelligent distributed systems and applications''.
\\
\textbf{Artifact.} \emph{EnergyTwin} code and experimental scenarios are available at \url{https://github.com/mvishiu11/energy-twin}.

\bibliographystyle{splncs04}
\bibliography{references}

@inproceedings{lasseter2002microgrids,
  author    = {Robert H. Lasseter},
  title     = {MicroGrids},
  booktitle = {Proc. IEEE Power Engineering Society Winter Meeting},
  volume    = {1},
  pages     = {305--308},
  year      = {2002},
  doi       = {10.1109/PESW.2002.985003}
}

@article{guerrero2011hierarchical,
  author  = {Josep M. Guerrero and Juan C. Vasquez and Jose Matas and Luis G. de Vicu{\~n}a and Miguel Castilla},
  title   = {Hierarchical Control of Droop-Controlled AC and {DC} Microgrids---A General Approach Toward Standardization},
  journal = {IEEE Transactions on Industrial Electronics},
  volume  = {58},
  number  = {1},
  pages   = {158--172},
  year    = {2011},
  doi     = {10.1109/TIE.2010.2066534}
}

@techreport{sotos2015scope2,
  author       = {Sotos, Mary Elizabeth},
  title        = {{GHG Protocol Scope 2 Guidance: An Amendment to the GHG Protocol Corporate Standard}},
  institution  = {World Resources Institute (WRI) and World Business Council for Sustainable Development (WBCSD)},
  year         = {2015},
  address      = {Washington, DC},
  url          = {https://ghgprotocol.org/scope-2-guidance},
  note         = {Accessed: 2025-11-17}
}

@article{olivares2014trends,
  author  = {Daniel E. Olivares and Ali Mehrizi{-}Sani and Amir H. Etemadi and Claudio A. C{\~a}nizares and Reza Iravani and Mehrdad Kazerani and Amir H. Hajimiragha and Oriol Gomis{-}Bellmunt and Maryam Saeedifard and Rodrigo Palma{-}Behnke and Guillermo A. Jim{\'e}nez{-}Est{\'e}vez and Nikos D. Hatziargyriou},
  title   = {Trends in Microgrid Control},
  journal = {IEEE Transactions on Smart Grid},
  volume  = {5},
  number  = {4},
  pages   = {1905--1919},
  year    = {2014},
  doi     = {10.1109/TSG.2013.2295514}
}

@techreport{king2004sandia,
  author      = {David L. King and William E. Boyson and Jay A. Kratochvil},
  title       = {Photovoltaic Array Performance Model},
  institution = {Sandia National Laboratories},
  number      = {SAND2004-3535},
  year        = {2004},
  doi         = {10.2172/919131}
}

@article{desoto2006pv,
  author  = {W. De Soto and S. A. Klein and W. A. Beckman},
  title   = {Improvement and Validation of a Model for Photovoltaic Array Performance},
  journal = {Solar Energy},
  volume  = {80},
  number  = {1},
  pages   = {78--88},
  year    = {2006},
  doi     = {10.1016/j.solener.2005.06.010}
}

@article{perez1987diffuse,
  author  = {Richard Perez and Robert Seals and Pierre Ineichen},
  title   = {A New Simplified Version of the {Perez} Diffuse Irradiance Model for Tilted Surfaces},
  journal = {Solar Energy},
  volume  = {39},
  number  = {3},
  pages   = {221--231},
  year    = {1987},
  doi     = {10.1016/S0038-092X(87)80031-2}
}

@article{ineichen2002linke,
  author  = {Pierre Ineichen and Richard Perez},
  title   = {A New Airmass Independent Formulation for the Linke Turbidity Coefficient},
  journal = {Solar Energy},
  volume  = {73},
  number  = {3},
  pages   = {151--157},
  year    = {2002},
  doi     = {10.1016/S0038-092X(02)00029-0}
}

@article{holmgren2018pvlib,
  author  = {William F. Holmgren and Clifford W. Hansen and Mark A. Mikofski},
  title   = {\texttt{pvlib} {Python}: A {Python} Package for Modeling Solar Energy Systems},
  journal = {Journal of Open Source Software},
  volume  = {3},
  number  = {29},
  pages   = {884},
  year    = {2018},
  doi     = {10.21105/joss.00884}
}

@article{chen2006battery,
  author  = {Ming Chen and Gabriel A. Rinc{\'o}n-Mora},
  title   = {Accurate Electrical Battery Model Capable of Predicting Runtime and {I--V} Performance},
  journal = {IEEE Transactions on Energy Conversion},
  volume  = {21},
  number  = {2},
  pages   = {504--511},
  year    = {2006},
  doi     = {10.1109/TEC.2006.874229}
}

@book{plett2015bms,
  author    = {Gregory L. Plett},
  title     = {Battery Management Systems, Volume I: Battery Modeling},
  publisher = {Artech House},
  year      = {2015},
  address   = {Norwood, MA}
}

@article{hong2016stlf,
  author  = {Tao Hong and Shu Fan},
  title   = {Probabilistic Electric Load Forecasting: A Tutorial Review},
  journal = {International Journal of Forecasting},
  volume  = {32},
  number  = {3},
  pages   = {914--938},
  year    = {2016},
  doi     = {10.1016/j.ijforecast.2015.11.011}
}

@article{antonanzas2016solar,
  author  = {J. Antonanzas and N. Osorio and R. Escobar and R. Urraca and F. J. Martinez-de-Pison and F. Antonanzas-Torres},
  title   = {Review of Photovoltaic Power Forecasting},
  journal = {Renewable and Sustainable Energy Reviews},
  volume  = {38},
  pages   = {587--607},
  year    = {2016},
  doi     = {10.1016/j.rser.2015.12.145}
}

@article{hu2021mpcoverview,
  author  = {Jian Hu and Wen Shi and Miroslav Krstic and Josep M. Guerrero},
  title   = {Model Predictive Control of Microgrids: An Overview},
  journal = {Renewable and Sustainable Energy Reviews},
  volume  = {134},
  pages   = {110301},
  year    = {2021},
  doi     = {10.1016/j.rser.2020.110301}
}

@article{erseghe2014admmopf,
  author  = {Tomaso Erseghe},
  title   = {Distributed Optimal Power Flow Using {ADMM}},
  journal = {IEEE Transactions on Power Systems},
  volume  = {29},
  number  = {5},
  pages   = {2370--2380},
  year    = {2014},
  doi     = {10.1109/TPWRS.2014.2306495}
}

@techreport{fipa2002cnp,
  author      = {{FIPA}},
  title       = {FIPA Contract Net Interaction Protocol Specification},
  institution = {Foundation for Intelligent Physical Agents},
  number      = {SC00029H},
  year        = {2002},
  url         = {https://www.fipa.org/specs/fipa00029/SC00029H.html}
}

@misc{energytwin_github2025,
  author    = {mvishiu11 and contributors},
  title     = {EnergyTwin: Agentic System for Energy Microgrids (code repository)},
  year      = {2025},
  howpublished = {\url{https://github.com/mvishiu11/energy-twin}},
  note      = {commit/version accessed Oct. 2, 2025}
}

@techreport{certs2002whitepaper,
  author      = {Robert H. Lasseter and Abbas Akhil and Chris Marnay and John Stephens and Jeff Dagle and Ross Guttromson and A. Sakis Meliopoulos and Robert Yinger and Joe Eto},
  title       = {Integration of Distributed Energy Resources: The {CERTS} Microgrid Concept},
  institution = {U.S. Department of Energy},
  year        = {2002},
  doi         = {10.2172/799644},
  url         = {https://eta-publications.lbl.gov/sites/default/files/certs_microgrid_concept_final.pdf}
}

@article{uddin2023microgrids,
  author  = {Moslem Uddin and Huadong Mo and Daoyi Dong and Sondoss Elsawah and Jianguo Zhu and Josep M. Guerrero},
  title   = {Microgrids: A Review, Outstanding Issues and Future Trends},
  journal = {Energy Strategy Reviews},
  year    = {2023},
  volume  = {49},
  pages   = {101127},
  doi     = {10.1016/j.esr.2023.101127}
}

@article{lara2024revisiting,
  author  = {Jos{\'e} Daniel Lara and Rodrigo Henriquez{-}Auba and Deepak Ramasubramanian and Sairaj V. Dhople and Duncan S. Callaway and Seth R. Sanders},
  title   = {Revisiting Power Systems Time-Domain Simulation Methods and Models},
  journal = {IEEE Transactions on Power Systems},
  year    = {2024},
  volume  = {39},
  number  = {2},
  pages   = {2421--2437},
  doi     = {10.1109/TPWRS.2023.3303291}
}

@article{thurner2018pandapower,
  author  = {Leon Thurner and Adil Malik and Manuel Scheidler and Fabian Schafer and Jens H. Menke and Juliane Dollichon and Friederike Meier and Steffen Meinecke and Martin Braun},
  title   = {pandapower---An Open-Source Python Tool for Convenient Modeling, Analysis, and Optimization of Electric Power Systems},
  journal = {IEEE Transactions on Power Systems},
  year    = {2018},
  volume  = {33},
  number  = {6},
  pages   = {6510--6521},
  doi     = {10.1109/TPWRS.2018.2829021}
}

@article{hardy2024helics,
  author  = {Todd D. Hardy and coauthors},
  title   = {{HELICS}: A Co-Simulation Framework for Scalable Multi-Domain Modeling and Analysis},
  journal = {IEEE Access},
  year    = {2024},
  volume  = {12},
  pages   = {24325--24345},
  note    = {Open-access version: NREL Report \#87610},
  url     = {https://docs.nrel.gov/docs/fy24osti/87610.pdf}
}

@article{barbierato2022cosim,
  author  = {Luca Barbierato and Pietro Rando Mazzarino and Marco Montarolo and Alberto Macii and Edoardo Patti and Lorenzo Bottaccioli},
  title   = {A Comparison Study of Co-Simulation Frameworks for Multi-Energy Systems: The Scalability Problem},
  journal = {Energy Informatics},
  year    = {2022},
  volume  = {5},
  number  = {Suppl 4},
  pages   = {53},
  doi     = {10.1186/s42162-022-00231-6}
}

@article{li2023hierarchical,
title = {Hierarchical Control for Microgrids: A Survey on Classical Methods and Machine Learning},
author = {Li, Zhen and O{~n}ate, Pedro and Tang, Yu and Yan, Zuofan and Guerrero, Josep M.},
journal = {Sustainability},
volume = {15},
number = {11},
pages = {8952},
year = {2023},
doi = {10.3390/su15118952}
}

@article{ahmed2023review,
title = {Review on Microgrids Design and Monitoring Approaches},
author = {Ahmed, Ilyas and Boudjadar, Jalila and Uz-Zaman, Zahid and others},
journal = {Scientific Reports},
volume = {13},
number = {1},
pages = {19916},
year = {2023},
doi = {10.1038/s41598-023-46888-9}
}

@article{aghazadeh2024dtses,
title = {Digital Twins of Smart Energy Systems: A Systematic Literature Review on Enablers, Design, Management and Computational Challenges},
author = {Aghazadeh~Ardebili, Ali and Zappatore, Marco and Ramadan, Amro~I.~H.~A. and Longo, Antonella and Ficarella, Antonio},
journal = {Energy Informatics},
volume = {7},
pages = {94},
year = {2024},
doi = {10.1186/s42162-024-00385-5}
}

@article{kumari2023dtmicrogrid,
title = {A Comprehensive Review of Digital Twin Technology for Grid-Connected Microgrid Systems: State of the Art, Potential and Challenges Faced},
author = {Kumari, Namita and Sharma, Ankush and Tran, Binh and Chilamkurti, Naveen and Alahakoon, Damminda},
journal = {Energies},
volume = {16},
number = {14},
pages = {5525},
year = {2023},
doi = {10.3390/en16145525}
}

@article{joshal2023mpc,
title = {Microgrids with Model Predictive Control: A Critical Review},
author = {Joshal, Keshav Singh and Gupta, Akshay},
journal = {Energies},
volume = {16},
number = {13},
pages = {4851},
year = {2023},
doi = {10.3390/en16134851}
}

@inproceedings{bellifemine2007jade,
title = {Developing Multi-Agent Systems with {JADE}},
author = {Bellifemine, Fabio and Caire, Giovanni and Greenwood, Dominic},
booktitle = {Agent Oriented Software Engineering},
publisher = {Wiley},
year = {2007}
}

@article{izmirlioglu2024massurvey,
title = {A Survey of Multi-Agent Systems for Smartgrids},
author = {Izmirlioglu, Yusuf and Pham, Loc and Son, Tran Cao and Pontelli, Enrico},
journal = {Energies},
volume = {17},
number = {15},
pages = {3620},
year = {2024},
doi = {10.3390/en17153620}
}

@article{elhafiane2024mas,
title = {A Multi-Agent System Approach for Real-Time Energy Management and Control in Hybrid Low-Voltage Microgrids},
author = {El Hafiane, Dounia and Essaaidi, Mohsine and Hmina, Nouredine and others},
journal = {Energy AI},
volume = {16},
pages = {100431},
year = {2024},
doi = {10.1016/j.egyai.2024.100431}
}

@article{dugan2011opendss,
title = {{OpenDSS}: The Open Distribution System Simulator},
author = {Dugan, Roger C. and Montenegro, Daniel},
journal = {IEEE PES General Meeting (tutorial/notes)},
year = {2011},
note = {EPRI OpenDSS Reference; widely used for distribution system quasi-static time series},
url = {https://smartgrid.epri.com/SimulationTool.aspx}
}

@article{brown2018pypsa,
title = {PyPSA: Python for Power System Analysis},
author = {Brown, Tom and H{"o}rsch, Jonas and Schlachtberger, David},
journal = {Journal of Open Research Software},
volume = {6},
pages = {4},
year = {2018},
doi = {10.5334/jors.188}
}

@article{chassin2014gridlabd,
title = {GridLAB-D: An Agent-based Simulation Framework for Smart Grids},
author = {Chassin, David P. and Fuller, Jason C. and Djilali, Ned and Earle, Luke},
journal = {arXiv preprint},
eprint = {1405.3136},
year = {2014}
}

@misc{su2024opensourcedynamic,
title = {A Survey of Open-Source Power System Dynamic Simulators with Grid-Forming Inverter for Machine Learning Applications},
author = {Su, Tong and Peng, Jiangkai and Selim, Alaa and Zhao, Junbo and Tan, Jin},
howpublished = {arXiv:2412.08065},
year = {2024},
url = {https://arxiv.org/abs/2412.08065}

}

@article{machele2024interconnected,
title = {Interconnected Smart Transactive Microgrids—Review and Trends},
author = {Machele, Givais D. and Anvari-Moghaddam, Amjad and Naghavi, Nafiseh and others},
journal = {Electronics},
volume = {13},
number = {10},
pages = {1949},
year = {2024},
doi = {10.3390/electronics13101949}
}

@article{bahrani2024ibrlandscape,
  title        = {Grid-Forming Inverter-Based Resource Research Landscape: Understanding the Key Assets for Renewable-Rich Power Systems},
  author       = {Bahrani, Behrooz and Ravanji, Mohammad Hasan and Kroposki, Benjamin},
  journal      = {IEEE Power \& Energy Magazine},
  volume       = {22},
  number       = {2},
  pages        = {18--29},
  year         = {2024},
  doi          = {10.1109/MPE.2023.3343338}
}

@techreport{esig2024gfm,
  title        = {Grid-Forming Technology in Energy Systems Integration},
  author       = {{Energy Systems Integration Group}},
  year         = {2024},
  note         = {Technical Report},
  url          = {https://www.esig.energy/wp-content/uploads/2024/10/ESIG-GFM-report-2022.pdf}
}

@article{breiman2001rf,
  title={Random forests},
  author={Breiman, Leo},
  journal={Machine learning},
  volume={45},
  number={1},
  pages={5--32},
  year={2001},
  publisher={Springer}
}

@misc{ortools2024,
  title={{Google OR-Tools}},
  author={Google},
  year={2024},
  howpublished={\url{https://developers.google.com/optimization}}
}
\end{document}